\documentclass[linenumbers=off]{aastex631}

\graphicspath{{./}{Figures/}}

\definecolor{dgreen}{rgb}{0.1, 0.53, 0.22}

\usepackage{ulem}
\usepackage{CJK}
\usepackage{hyperref}
\usepackage{amsmath}
\usepackage{graphicx}
\usepackage{multirow}
\usepackage{booktabs} 
\usepackage{threeparttable} 

\begin{document}

\title{Phase II of the LAMOST-\textit{Kepler}/\textit{K2} Survey. II. Time Domain of Medium-resolution Spectroscopic Observations from 2018 to 2023}

\author[0000-0002-0040-8351]{Mingfeng Qin}
\affiliation{Institute for Frontiers in Astronomy and Astrophysics, Beijing Normal University, Beijing 102206, P.~R.~China}
\affiliation{School of Physics and Astronomy, Beijing Normal University, Beijing 100875, P.~R.~China}

\author[0000-0001-8241-1740]{Jian-Ning Fu}
\affiliation{Institute for Frontiers in Astronomy and Astrophysics, Beijing Normal University, Beijing 102206, P.~R.~China}
\affiliation{School of Physics and Astronomy, Beijing Normal University, Beijing 100875, P.~R.~China}

\author[0000-0002-7660-9803]{Weikai Zong}
\affiliation{Institute for Frontiers in Astronomy and Astrophysics, Beijing Normal University, Beijing 102206, P.~R.~China}
\affiliation{School of Physics and Astronomy, Beijing Normal University, Beijing 100875, P.~R.~China}

\author[0000-0001-5419-2042]{Peter De Cat}
\affiliation{Royal Observatory of Belgium, Ringlaan 3, B-1180 Brussel, Belgium}

\author[0000-0002-0474-0896]{Antonio Frasca}
\affil{INAF -- Osservatorio Astrofisico di Catania, Via S. Sofia 78, I-95123 Catania, Italy}

\author[0000-0003-3816-7335]{Tianqi Cang}
\affiliation{Institute for Frontiers in Astronomy and Astrophysics, Beijing Normal University, Beijing 102206, P.~R.~China}
\affiliation{School of Physics and Astronomy, Beijing Normal University, Beijing 100875, P.~R.~China}

\author{Jiangtao Wang}
\affil{Department of Astronomy, Beijing Normal University, Beijing~100875, P.~R.~China}

\author{Jianrong Shi}
\affil{Key Lab for Optical Astronomy, National Astronomical Observatories, Chinese Academy of Sciences, Beijing 100012, P.~R.~China}

\author{Ali Luo}
\affil{Key Lab for Optical Astronomy, National Astronomical Observatories, Chinese Academy of Sciences, Beijing 100012, P.~R.~China}

\author{Haotong Zhang}
\affil{Key Lab for Optical Astronomy, National Astronomical Observatories, Chinese Academy of Sciences, Beijing 100012, P.~R.~China}

\author{Hongliang Yan}
\affil{Key Lab for Optical Astronomy, National Astronomical Observatories, Chinese Academy of Sciences, Beijing 100012, P.~R.~China}

\author{J. Molenda- \.Zakowicz}
\affil{Astronomical Institute of the University of Wroc\l{}aw, ul. Kopernika 11, 51-622 Wroc\l{}aw, Poland}

\author{R. O. Gray}
\affil{Department of Physics and Astronomy, Appalachian State University, Boone, NC 28608, USA}

\author{Jiaxin Wang}
\affil{Department of Astronomy, Beijing Normal University, Beijing~100875, P.~R.~China}

\correspondingauthor{Jian-Ning Fu}
\email{jnfu@bnu.edu.cn}

\begin{abstract} 

The LAMOST-\textit{Kepler}/\textit{K2} Medium-Resolution Spectroscopic Survey (LK-MRS) conducted time-domain medium-resolution spectroscopic observations of 20 LAMOST plates in the \textit{Kepler} and \textit{K2} fields from 2018 to 2023, a phase designated as LK-MRS-\uppercase\expandafter{\romannumeral1}. A catalog of stellar parameters for a total of 36,588 stars, derived from the spectra collected during these five years, including the effective temperature, the surface gravity, the metallicity, the $\alpha$-element abundance, the radial velocity, and $v \sin i$ of the target stars, is released, together with the weighted averages and uncertainties. At S/N = 10, the measurement uncertainties are $120\,\mathrm{K}$, $0.18\,\mathrm{dex}$, $0.13\,\mathrm{dex}$, $0.08\,\mathrm{dex}$, $1.9\,\mathrm{km/s}$, and $4.0\,\mathrm{km/s}$ for the above parameters, respectively. Comparisons with the parameters provided by the APOGEE and GALAH surveys validate the effective temperature and surface gravity measurements, showing minor discrepancies in metallicity and $\alpha$-element abundance values. We identified some peculiar star candidates, including 764 metal-poor stars, 174 very metal-poor stars, and 30 high-velocity stars. Moreover, we found 2,333 stars whose radial velocity seems to be variable. Using \textit{Kepler}/\textit{K2} or \textit{TESS} photometric data, we confirmed 371 periodic variable stars among the radial velocity variable candidates and classified their variability types. LK-MRS-\uppercase\expandafter{\romannumeral1} provides spectroscopic data being useful for studies of the \textit{Kepler} and \textit{K2} fields. The LK-MRS project will continue collecting time-domain medium-resolution spectra for target stars during the third phase of LAMOST surveys, providing data to support further scientific research.
\end{abstract}


\section{Introduction}\label{sec:intro}

The \textit{Kepler} satellite, launched in 2009 and operational until 2018, successfully carried out the \textit{Kepler} and \textit{K2} missions successively, which involved continuous photometric monitoring of more than 780,000 stars in total \citep{2010Borucki, 2014Howell}. The high-precision photometric data obtained from \textit{Kepler}/\textit{K2} have been extensively utilized in diverse scientific domains, including exoplanet detection, stellar astrophysics, Galaxy studies, and investigations of solar system objects \citep{2018Barentsen}. Despite the important contributions of photometric data, a comprehensive understanding of stellar properties requires more than photometry alone. An accurate and homogeneous determination of atmospheric parameters, such as that achievable with large spectroscopic surveys, is invaluable for a thorough source characterization \citep{2020Worley} and to complement their highly precise light curves collected by space missions.

To this aim several ground-based spectroscopic observation programs have been developed, such as the APOKASC survey \citep{2014Pinsonneault,2017Serenelli,2018Pinsonneault}, the LAMOST-\textit{Kepler} project (hereafter LK-project) \citep{2015DeCat,2020Fu}, and the \textit{K2}-HERMES survey \citep{2018Wittenmyer}. These surveys aim to provide homogeneous spectroscopic follow-up observations of stars targeted by the \textit{Kepler} and \textit{K2} missions, utilizing multi-object spectrographs capable of observing numerous targets simultaneously. For example, APOKASC uses the high-resolution (R=22,500) H-band spectrograph with 230 fibers from APOGEE on the SDSS 2.5 m telescope to observe \textit{Kepler} targets \citep{2014Pinsonneault}. The LK-project utilizes the LAMOST telescope with an effective aperture of $\sim$ 4 meters and its 4000 fibers to perform low-resolution (R$\sim$1800) or medium-resolution (R$\sim$7500) spectroscopic observations of \textit{Kepler} and \textit{K2} fields \citep{2015DeCat,2018Zong,2020Fu,2020Wang}. Similarly, the \textit{K2}-HERMES survey employs the HERMES multi-object spectrograph on the 3.9 m Anglo-Australian Telescope, which delivers high-resolution spectra with a resolving power of $R \sim 28,000$ in a single exposure, enabling simultaneous observations of up to 360 stars \citep{2018Wittenmyer}.

The LK-project was launched in 2010 \citep{2015DeCat} with the primary goal of conducting low-resolution (R=1800, {\citeauthor{2012Zhao} \citeyear{2012Zhao}}) spectroscopic observations of the \textit{Kepler} and \textit{K2} fields. It began with the \textit{Kepler} field in 2012 (LK1, {\citeauthor{2015DeCat} \citeyear{2015DeCat}}) and later expanded to the \textit{K2} field in 2015 (LK2, {\citeauthor{2020Wang} \citeyear{2020Wang}}). The spectra and stellar parameters derived from the LK-project have been applied by astronomers in a wide range of research areas, including stellar parameter determination, stellar pulsations and asteroseismology, exoplanet studies, stellar magnetic activity and flares, peculiar stars in the Milky Way, and binary star systems \citep{2020Fu}. In order to obtain time-domain spectra for stars in the \textit{Kepler} and \textit{K2} fields, the LK-project continues alongside the launch of the LAMOST Phase II survey.

Since September 2017, the upgrade of LAMOST spectrographs have enhanced the telescope's capabilities with the ability of making medium-resolution (R=7500) spectroscopic observation surveys. This enables the pursuit of wide fields of scientific goals, including binary/multiple star systems, stellar pulsations, star formation, emission nebulae, galactic archaeology, exoplanet host stars, and open clusters \citep{2020Liu,2022Yan}. Subsequently, in October 2018, the LK Medium-Resolution Spectroscopic Survey (LK-MRS) was launched \citep[hereafter Paper\,1]{2020Zong}. This survey selected 20 observational plates of LAMOST for time-domain (TD) spectroscopic monitoring. A total of 12,860 stars are located in the \textit{Kepler} field, among which 9,262 have been previously observed by the \textit{Kepler} mission. In addition, 41,425 stars are located in the \textit{K2} field, among which 19,540 have been observed by the \textit{K2} mission. In total, 53.06\% of the targets have been previously observed by either the \textit{Kepler} or \textit{K2} missions. Paper\,1 presents a catalog of stellar parameters derived from one year of observation data, including weighted average values and their associated uncertainties. The study also evaluated the uncertainty in these parameters and discussed the scientific prospects in several key areas.

As discussed in Paper 1, the LK-MRS dataset has been extensively used since its launch. It has been particularly valuable in the study of multiple star systems, including starspot-modulated eclipsing binaries \citep{2020Pan,2024Pan,2024Wang}, spectroscopic binaries \citep{2022Zhang}, pulsating binaries \citep{2024Jin}, and ultracool dwarf comoving companions \citep{2024Rothermich}. The dataset has advanced asteroseismology, enhancing our understanding of stellar pulsations and internal structures in variable stars \citep{2023Ma,2021Wang,2024Zong}. LK-MRS data have also been crucial for investigating stellar activity, shedding light on the evolution of magnetic activity in solar-like stars \citep{2023Mathur}, chromospheric activities \citep{2023Han,2024Li}, and extreme stellar prominence eruptions \citep{2025Lu}. Furthermore, LK-MRS spectra have contributed to determining self-consistent stellar radial velocities \citep{2021Zhang}, characterizing Kepler targets using ROTFIT \citep{2022Frasca}, and estimating stellar parameters and chemical abundances through Cycle-StarNet \citep{2023Wang}. These findings not only demonstrate the significant scientific value of the LK-MRS dataset, but also emphasize the critical need for its sustained development. The continuing spectroscopic observations in the LK-MRS survey are particularly crucial, as the achieved time-domain spectroscopic coverage provides an essential foundation for in-depth investigations of variable astrophysical sources.

By June 2023, LAMOST had completed its second phase of sky surveys, and concurrently, LK-MRS concluded its five-year regular survey, referred to as LK-MRS-\uppercase\expandafter{\romannumeral1}. This paper presents the observational results of the LK-MRS-\uppercase\expandafter{\romannumeral1}. Section\,2 provides a comprehensive overview of the LK-MRS observing strategy, target selection, and the outcomes of LK-MRS-\uppercase\expandafter{\romannumeral1}. Section\,3 examines the atmospheric parameters derived from the LASP pipeline of LK-MRS-\uppercase\expandafter{\romannumeral1}, evaluating both internal and external errors. In Section\,4, we identify a number of candidates for peculiar stars, including high-velocity stars, metal-poor stars, and those exhibiting radial velocity variations. Finally, Section\,5 summarizes the key findings of this study.


\section{LK-MRS Observation Strategy and Five-Year Data}\label{sec:lk_mrs}

\subsection{Observation Strategy and Targets of the LK-MRS}

Since October 2018, LAMOST has launched its second-phase surveys, which include both low-resolution spectroscopic survey (LRS) and MRS. The MRS is primarily conducted around the full moon nights, accounting for approximately 50$\%$ of the observation time for each lunar month, and is divided into non-TD (NT) and TD modes. The NT MRS mode is designed for large-scale spectroscopic mapping. Each plate is observed once, with three consecutive 1200-second exposures per visit, and the resulting spectra are co-added to improve S/N. In contrast, the TD MRS mode aims to monitor spectral variability by repeatedly observing each plate over time. The target stars on each plate remain fixed, and each visit includes 3–8 single 1200-second exposures. The same stars are revisited across multiple nights throughout the five-year survey \citep{2020Liu}. Spectroscopic observations of TD MRS can provide atmospheric parameters for variable stars, facilitating more in-depth studies of these stars \citep{2022Li}. They are also of great help in the study of spectroscopic binaries, chromospheric activity variations, and flares \citep{2020Pan,2022Frasca,Frasca2025}. The TD MRS comprises four sub-projects, one of which is the LK-MRS \citep{2020Zong}.

\begin{figure*}
\centering
\includegraphics[scale=0.72]{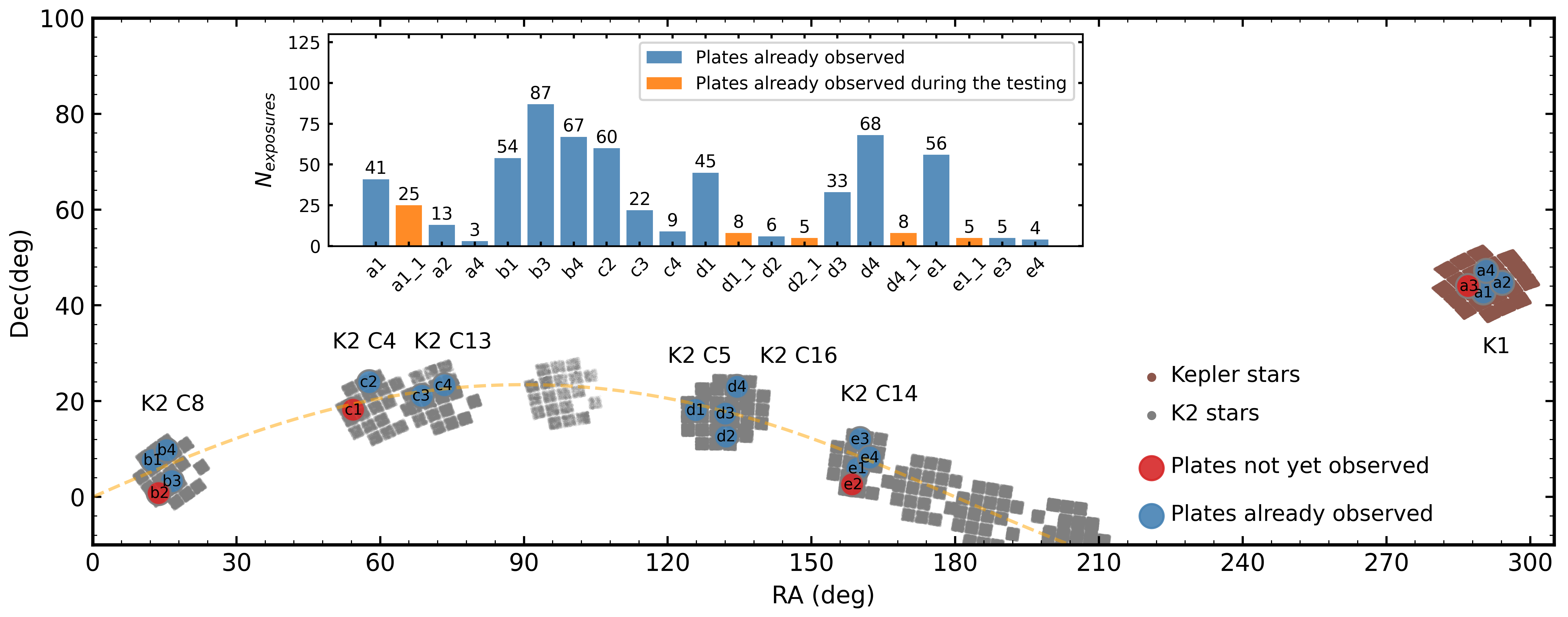}
\caption{Spatial distribution of plates of the LK–MRS-\uppercase\expandafter{\romannumeral1}. The yellow dash line represents the ecliptic plane. 
The inset shows the exposure counts for each plate. Blue bars represent plates that have been observed, while orange bars indicate plates observed during the testing phase.} \label{fig:observe}
\end{figure*}

To integrate \textit{Kepler} time-series photometry with TD MRS data and enable extensive large-sample studies of \textit{Kepler}/\textit{K2} targets, the LK-MRS has strategically selected 4 observational plates within the \textit{Kepler} field and 16 plates in the \textit{K2} fields, for a total of 20 plates \citep{2020Zong}. For each plate, the set of target stars is fixed, and repeated spectroscopic observations are scheduled to monitor variability. Each plate is planned for 60 visits, with their spatial distribution indicated by the blue and red circular footprints in Figure \ref{fig:observe}. 
All selected plates are positioned above a declination of $-10^\circ$ to ensure that they remain within LAMOST's observable ranges. The limiting magnitude is set at $G = 15.0$ mag, making these stars viable for observation during full moon nights. For a comprehensive description of the target selection methods applied in the LK-MRS, see Paper\,1.

\subsection{Data of the LK-MRS-\uppercase\expandafter{\romannumeral1}}

\begin{table*}[t]
\centering
\caption{Data Overview of the LK-MRS-\uppercase\expandafter{\romannumeral1}.}
\label{tab:yearly_observations}
\begin{threeparttable}
\begin{tabular*}{\textwidth}{@{\extracolsep{\fill}} c c c r r r r r r r}
\toprule
  Year & $N_{\text{Plate}}$ & $N_{\text{Exposures}}$ & $N_{\text{Spectra}}$ & $N_{T_{\text{eff}}}$ & $N_{\log g}$ & $N_{[\text{Fe}/\text{H}]}$ & $N_{\text{RV}}$ & $N_{v \sin i}$ & $N_{[\alpha/\text{M}]}$ \\
\midrule
 2018 & 25 & 136 & 791,779 & 34,211 & 34,211 & 34,211 & 32,377 & 4,557 & 18,572 \\
 2019 & 40 & 201 & 1,111,398 & 48,905 & 48,905 & 48,905 & 46,199 & 6,412 & 23,846 \\
 2020 & 34 & 142 & 865,996 & 48,263 & 48,263 & 48,263 & 45,619 & 7,360 & 27,309 \\
 2021 & 14 & 63 & 357,414 & 17,531 & 17,531 & 17,531 & 16,454 & 2,525 & 11,012 \\
 2022 & 11 & 41 & 191,438 & 12,679 & 12,679 & 12,679 & 12,001 & 1,628 & 7,892 \\
 2023 & 10 & 41 & 213,760 & 9,048 & 9,048 & 9,048 & 8,620 & 1,328 & 5,444 \\
 Total & 134 & 624 & 3,531,785 & 170,637 & 170,637 & 170,637 & 161,270 & 23,810 & 94,075 \\
\bottomrule
\end{tabular*}
\begin{tablenotes}
\item Column\,1 represents the year of observation, Column\,2 shows the number of plates observed each year, Column\,3 displays the number of visits per plate annually, Column\,4 indicates the total spectra observed each year, Column\,5 to 10 denote the count of stellar parameters derived using LASP annually.
\end{tablenotes}
\end{threeparttable}
\end{table*}

\begin{figure*}
\centering
\includegraphics[scale=0.62]{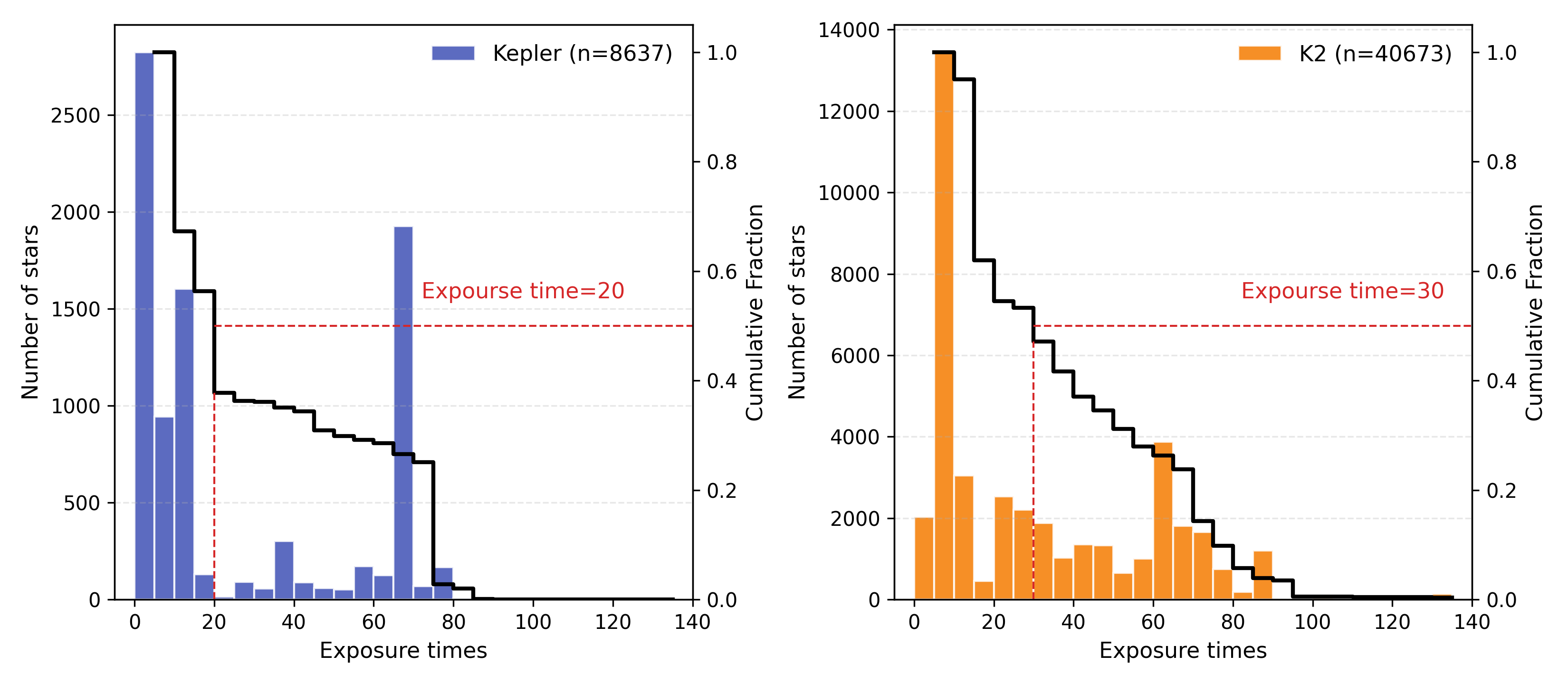}
\caption{Distribution of exposure times for LK-MRS-\uppercase\expandafter{\romannumeral1} sources in the \textit{Kepler} (left) and \textit{K2} (right) surveys. The histogram (vertical bars) shows stellar counts per exposure bin, while the stepped curve (black) displays the reverse cumulative distribution function (CDF) computed from the maximum exposure time downward (i.e., fraction of stars with exposure time $\geq t$). The red dashed line indicates the median exposure (50\textsuperscript{th} percentile), with its numerical value annotated in the plot.}\label{fig:exposure_times}
\end{figure*}

The LK-MRS-\uppercase\expandafter{\romannumeral1} conducted TD spectroscopic observations from September 2018 to June 2023, covering both the \textit{Kepler} and \textit{K2} fields. The distribution of all observed plates in the celestial coordinate system is shown in Figure \ref{fig:observe}. In this figure, the blue circular markers represent plates that have already been observed, while the red circular markers indicate plates that have not yet been observed. As of June 2023, only four plates remain unobserved, three located in the \textit{K2} fields and one in the \textit{Kepler} field. The inset in Figure \ref{fig:observe} presents a histogram of the exposure counts for each plate, with blue bars representing 16 of the plates planned for observation in Table\,1 of Paper\,1 and orange bars representing the five plates from the testing phase\footnote{The plan IDs for these plates are HIP95119KP01, TD082325N180811K02, TD084844N123545K02, TD085754N225914K02, and TD103827N055449K02, which correspond to the same fields as K1a1, K2d1, K2d2, K2e1, and K2d4, respectively. We refer to them as K1a1\_1, K2d1\_1, K2d2\_1, K2e1\_1, and K2d4\_1 in this paper.}. The inset shows that for four plates 60 or more observations have been made, while another four already have more than 40 observations. In total, 624 exposures were obtained, completing approximately 52\% of the planned observations. 

Each exposure produces two single-exposure spectra, one from the red arm and another one from the blue arm of the spectrograph. In most cases, three consecutive exposures of the same target are coadded to improve the signal-to-noise ratio (S/N), resulting in coadded red-arm and blue-arm spectra. LK-MRS-\uppercase\expandafter{\romannumeral1} yielded 3,531,785 spectra, including 2,941,929 single-exposure spectra and 589,856 coadded spectra. These spectra correspond to 49,310 unique stellar targets, comprising 8,637 stars in the \textit{Kepler} field and 40,673 in the \textit{K2} field. The distribution of the number of exposures per star is illustrated in Figure\,\ref{fig:exposure_times}. The blue-purple bars and orange represent the histograms of exposure counts for all stars, 8,637 stars in the \textit{Kepler} fields, and 40,673 stars in the \textit{K2} field, respectively. The cumulative distribution functions for all three samples are overplotted to facilitate comparison. In addition, the red dashed line indicates the average number of exposures, which is 20 and 30. The number of exposures in the \textit{Kepler} field is significantly lower than in the \textit{K2} field, mainly because the \textit{Kepler} observations fall in the summer season, during which LAMOST undergoes regular maintenance in July and August. Additionally, frequent rainfall during summer further reduces observational efficiency.

Among the 589,856 coadded spectra collected, half correspond to blue-arm coadded spectra and the other half to red-arm coadded spectra. Due to the limitations of the LAMOST Stellar Parameter Pipeline (LASP), which is capable of deriving stellar parameters only for late-A and FGK-type stars, stellar atmospheric parameters were successfully determined for 170,637 spectra. Their physical parameters, including effective temperature ($T_\text{eff}$), surface gravity ($\log g$), and metallicity ([Fe/H]), were derived using the latest version of the LASP pipeline\footnote{\url{https://www.lamost.org/lmusers/}}. A summary of annual observations, including the number of plates, exposures, collected spectra, and parameter statistics, is provided in Table\,\ref{tab:yearly_observations}.

This work presents a statistical analysis of LK-MRS based on the LAMOST DR11\footnote{\url{https://www.lamost.org/lmusers/} dataset, covering observations from September 2018 to June 2023. Compared to Paper\,I, which only included data from a limited time span (September 2018 to June 2019), the current study incorporates four additional years of observations. The early data included in Paper\,I have been fully reprocessed using the latest version of the LASP pipeline, resulting in improved calibration accuracy and enhanced consistency across the entire dataset. The current dataset includes 401 additional plate exposures and 2,366,133 more spectra relative to Paper\,I. These additions not only contribute new stellar targets but also provide new epochs for previously observed stars, significantly improving the temporal coverage for time-domain analysis.} Therefore, the current release offers both new detections and improved measurements relative to Paper\,I, enabling more robust statistical and variability studies.

\section{Stellar Atmospheric Parameters of LK-MRS-\uppercase\expandafter{\romannumeral1}}

\subsection{Atmospheric Parameter Catalog of LK-MRS-\uppercase\expandafter{\romannumeral1}}

The LAMOST Stellar Parameter Pipeline (LASP) is an automated pipeline that determines stellar atmospheric parameters,such as \(T_\text{eff}\), \(\log\,g\), [Fe/H], and RV,by matching observed spectra with ELODIE templates\footnote{\url{http://www.lamost.org/dr11/}}. During the LK-MRS-\uppercase\expandafter{\romannumeral1} survey, LASP derived stellar parameters from 170,637 coadded spectra. These spectra correspond to 36,588 unique stars. The difference arises because many stars were observed multiple times over the course of the survey to support time-domain analyses, and each valid coadded spectrum was processed independently. Specifically, 17,996 stars have at least three valid coadded spectra, enabling variability and rotational studies. In addition to the core LASP parameters, projected rotational velocities (\(v \sin i\)) were derived using PHOENIX templates \citep{2024Zuo}, limited to stars with \(T_\text{eff}\) between 5000 and 8500 K due to resolution constraints, resulting in 23,810 \(v \sin i\) measurements. However, we note that, as shown by \citet{2022Frasca}, \(v \sin i\) values smaller than 8 km/s cannot be reliably derived due to the resolution and sampling of MRS spectra, and only an upper limit of 8 km/s can be provided in these cases. 94,075 \([\alpha/M]\) measurements were derived in LK-MRS-\uppercase\expandafter{\romannumeral1}. Similarly, 161,270 RV measurements were obtained by matching the spectra with Kurucz templates \citep{2019Wang}.

\begin{table*}
\scriptsize
\centering
\caption{Stellar atmospheric parameters provided by LK-MRS-\uppercase\expandafter{\romannumeral1}.}
\label{tab:parameters}
\begin{threeparttable}
\begin{tabular*}{\textwidth}{p{0.05cm} p{2.0cm} p{1.1cm} p{1.1cm} p{1.1cm} p{1.1cm} p{0.8cm} p{1.0cm} p{1.1cm} p{1.0cm} p{1.0cm} p{1.6cm}}
\toprule
 \multicolumn{1}{c}{Times} & \multicolumn{1}{c}{uid} &{KIC/EPIC} & \multicolumn{1}{c}{RA} & \multicolumn{1}{c}{Dec} & \multicolumn{1}{c}{\(T_\text{eff}\)} & \multicolumn{1}{c}{$\log\,g$} & \multicolumn{1}{c}{$[\mathrm{Fe/H}]$} & {$[\alpha/\mathrm{M}]$} & \multicolumn{1}{c}{RV} & \multicolumn{1}{c}{$v\,\sin\,i$}& {label} \\
 \multicolumn{1}{c}{} & \multicolumn{1}{c}{} & \multicolumn{1}{c}{} & \multicolumn{1}{c}{(deg)} & \multicolumn{1}{c}{(deg)} & \multicolumn{1}{c}{(K)} & \multicolumn{1}{c}{(dex)} & \multicolumn{1}{c}{(dex)} & {(dex)} & \multicolumn{1}{c}{(km/s)} & \multicolumn{1}{c}{(km/s)}& \\
\midrule
12 & G17096966435490 & 220550719 & 10.580750 & 7.910431 & 5645.88 & 4.31 & -0.04 & 0.09 & -2.70 & -- & -- \\
   &                &                 &           &          & $\pm$ 39.79 & $\pm$ 0.06 & $\pm$ 0.03 & $\pm$ 0.02 & $\pm$ 0.41 & -- & -- \\
1  & G16297522311478 & 211440152 & 130.672740 & 12.248783 & 6477.13 & 4.12 & -0.09 & -- & 12.54 & -- & -- \\
   &                &                 &           &          & -- & -- & -- & -- & -- & -- & -- \\
1  & G16342182680148 & 212019272& 134.722330 & 20.573745 & 5704.37 & 4.03 & -0.20 & -- & 49.22 & 46.20 & -- \\
   &                &                 &           &          & -- & -- & -- & -- & -- & -- & -- \\
1  & G16346217356100 & 211524166& 131.273770 & 13.479998 & 5540.67 & 3.80 & -0.51 & 0.20 & 40.83 & -- & -- \\
   &                &                 &           &          & -- & -- & -- & -- & -- & -- & -- \\
6  & G16353924667769 & 211833343 & 133.371240 & 17.788899 & 4784.41 & 2.87 & -0.34 & 0.07 & 25.87 & -- & -- \\
   &                &                 &           &          & $\pm$ 12.74 & $\pm$ 0.03 & $\pm$ 0.00 & $\pm$ 0.01 & $\pm$ 0.97 & -- & -- \\
2  & L16715670508803 & 247247018  & 69.458066 & 20.067119 & 4675.25 & 4.63 & -0.67 & -- & 79.24 & -- & -- \\
   &                &                 &           &          & $\pm$ 0.07 & $\pm$ 0.01 & $\pm$ 0.02 & -- & $\pm$ 0.03 & -- & -- \\
11 & G17066013434593 & 220491346  & 12.628243 & 6.568253 & 5058.15 & 3.74 & -0.18 & 0.20 & -14.92 & 33.64 & -- \\
   &                &                 &           &          & $\pm$ 10.18 & $\pm$ 0.02 & $\pm$ 0.01 & $\pm$ 0.02 & $\pm$ 0.53 & $\pm$ 3.30 & -- \\
3  & L16700197857117 & 210867214  & 66.611719 & 20.886968 & 6094.06 & 4.29 & -0.57 & 0.15 & 7.52 & -- & -- \\
   &                &                 &           &          & $\pm$ 14.52 & $\pm$ 0.03 & $\pm$ 0.08 & $\pm$ 0.07 & $\pm$ 0.54 & -- & -- \\
9  & G16342062317845 & 212066917  & 133.435470 & 21.399934 & 5915.86 & 4.20 & 0.00 & -0.05 & -19.21 & -- & RVV \\
   &                &                 &           &          & $\pm$ 122.62 & $\pm$ 0.15 & $\pm$ 0.03 & $\pm$ 0.02 & $\pm$ 20.20 & -- &   \\
1  & G16340116868661 & 211953970 & 132.402140 & 19.532971 & 5283.99 & 4.10 & -0.58 & -- & 31.29 & -- & -- \\
   &                &                 &           &          & -- & -- & -- & -- & -- & -- & -- \\
... & ... &... & ... & ... &... &... &... &... &... &... &... \\
\bottomrule
\end{tabular*}
\end{threeparttable}
\begin{tablenotes}
\item Note: Column\,1 provides the number of available spectral parameters for each target, while Columns\,2 and 3 provide the uid of LAMOST and a combined identifier (EPIC/KIC), respectively. Columns\,4 and 5 list the RA and Dec of the stars in J2000.0. Columns\,6 to 11 present the stellar parameters, including \(T_\text{eff}\), $\log\,g$, [Fe/H], [$\alpha$/M], RV, and $v\,\sin\,i$, with their corresponding uncertainties listed below each value. The final column includes classifications or additional comments about the stars. Missing values are denoted by "–". The complete table will be accessible online in a machine-readable format.
\end{tablenotes}
\end{table*}

To estimate each stellar parameter, we used Equation\,1 from Paper\,1 to compute a weighted mean across multiple observations, yielding the final value,

\begin{equation}
    \overline{P} = \frac{\sum_k w_k \cdot P_k}{\sum_k w_k},
\end{equation}
where \( k \in [1, N] \) denotes the observation index for parameter \( P \) of an individual star. The weighting factor \( w_k \) is calculated based on the square of the S/N for each analyzed spectrum, thus giving greater importance to the spectra with higher S/N. We also applied Equation\,2 from Paper\,1 to calculate the weighted standard error for each parameter,

\begin{equation}
    \sigma_w (\overline{P}) = \sqrt{\frac{N}{N - 1} \cdot \frac{\sum_k w_k \cdot (P_k - \overline{P})^2}{\sum_k w_k}},
\end{equation}
where \( \sigma_w (\overline{P}) \) represents the standard error of the weighted mean \( \overline{P} \).
The final parameter values, their standard errors, and the number of observations for each star are summarized in Table \ref{tab:parameters}. The methodology for computing the mean stellar parameters follows exactly the same approach as described in Paper\,I. However, all spectra used in this work have been reprocessed using the updated LASP pipeline from LAMOST DR11, which incorporates improvements in calibration and parameter estimation. Moreover, the dataset used in this study includes four additional years of observations compared to Paper\,I, resulting in more spectra and higher observation counts for many stars. As a result, while the weighted mean parameters are generally consistent with those in Paper\,I, small differences may arise due to the inclusion of new measurements and the updated pipeline processing.

Table \ref{tab:parameters} includes information for a total of 36,588 stars, of which 18,892 stars are listed in the \textit{Kepler}/\textit{K2} input catalogs\citep{2011kepler,2016kepler,2016keplerlcsc,2016k2lightcurves}, accounting for approximately 51.6$\%$ of the total. The first column indicates the number of observation times. The second column lists the unique observation IDs (uid) assigned by LAMOST, while the third column provides the corresponding IDs from the \textit{Kepler} or \textit{K2} input catalogs, matched using a cross-matching radius of 3$\farcs$7\footnote{The radius of 3$\farcs$7 was adopted on the basis of the fiber pointing precision (0$\farcs$4) and the 3$\farcs$3 diameter of the fiber \citep{2018Zong}.}. In cases where multiple \textit{Kepler}/\textit{K2} targets fall within this matching radius, we selected the closest counterpart as the optimal match. IDs greater than 201,000,000 indicate targets from the \textit{K2} input catalog, while those less than 201,000,000 belong to the \textit{Kepler} input catalog. Blank entries indicate stars not included in either catalog. The fourth and fifth columns present the right ascension (RA) and declination (Dec) of the stars in J2000.0. Columns\,6 to 11 list the stellar parameters, including \(T_\text{eff}\), \(\log g\), [Fe/H], [\(\alpha\)/M], RV, and \(v \sin i\), with uncertainties provided below each value. The final column includes classifications or additional comments about the stars, further detailed in Section \ref{sec:discussion}. Missing values or parameters not determined by LASP are denoted as ``--''. The entire table will be published in machine-readable format as part of the online supplementary materials.

\subsection{Measurement uncertainties of the parameters of LK-MRS-\uppercase\expandafter{\romannumeral1}}

\begin{figure*}
\centering
\includegraphics[scale=0.7]{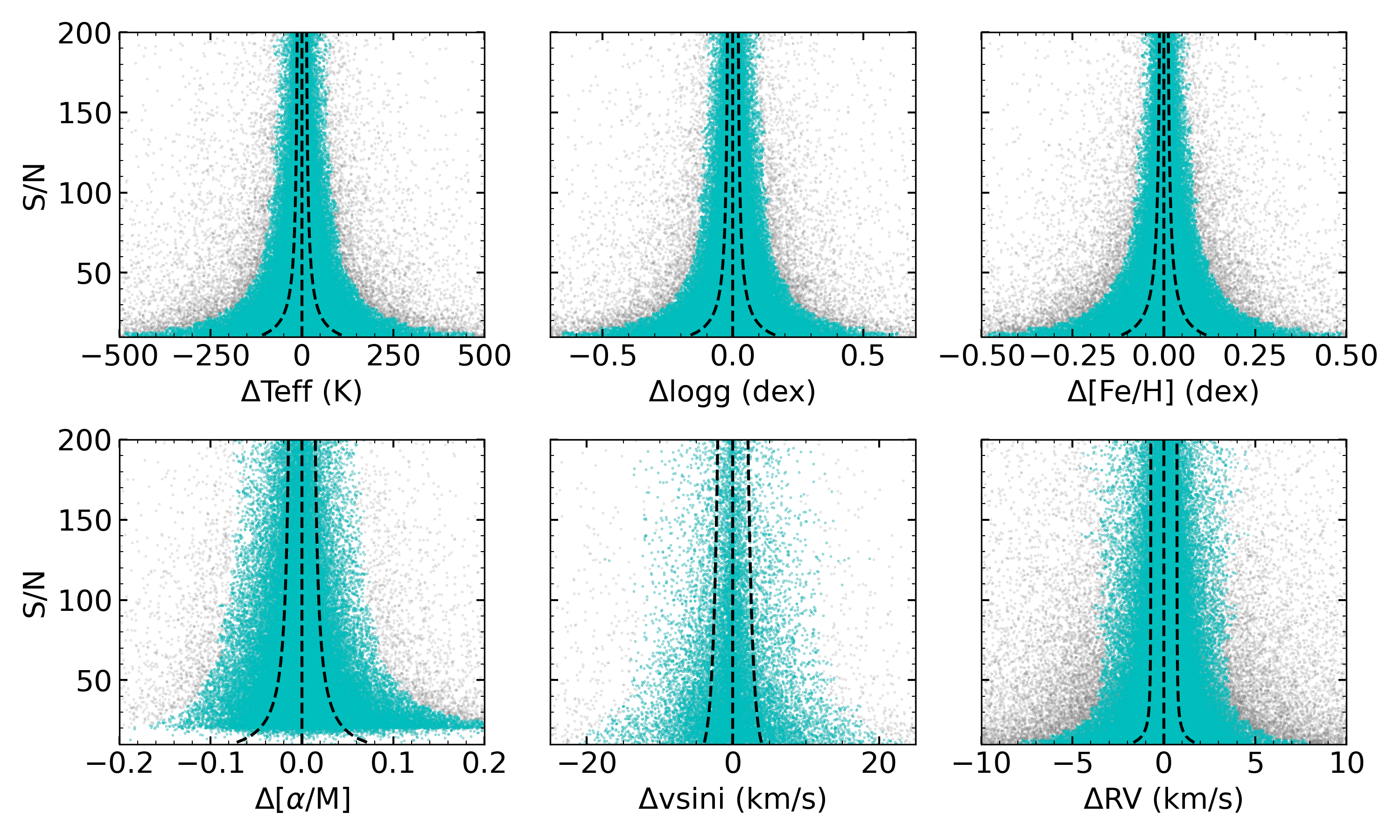}
\caption{S/N versus internal uncertainties for the five parameters \(T_\text{eff}\), $log\,g$, $[\alpha/M]$, RV, and \(v \sin i\). Gray points represent data identified as outliers, while cyan points correspond to data retained after filtering. The black dashed vertical line indicates the zero point on the x-axis, and the black dashed curves on either side represent the fitted relationship between \( \Delta P \) and S/N, as modeled by Equation\,4.} \label{fig:snr_delta}
\end{figure*}

For stars with multiple spectra measured by LASP, we conducted an assessment of their measurement uncertainties. This was achieved by calculating the difference between the parameters provided by each individual spectrum and the corresponding weighted average parameters, denoted as \( \Delta P \), using Equation\,3 from Paper\,1:

\begin{equation}
    \Delta P_k = (P_k - \overline{P}) \cdot \sqrt{\frac{N}{N - 1}},
\end{equation}
where \( \Delta P_k \) represents the deviation of each individual parameter \( P_k \) from the weighted average \( \overline{P} \), and \( N \) is the number of observations for that star. These differences serve as a measure of the internal uncertainty of each parameter. To visualize the relationship between \( \Delta P \) and the S/N, we plotted scatter diagrams for each parameter, as shown in Figure \ref{fig:snr_delta}. According to Figure \ref{fig:snr_delta}, the uncertainty of parameters reduce with S/N increasing. At S/N > 50 the scatter tends to settle towards a constant value and the errors on all the parameters are small enough for assessing a high data quality.

To minimize the impact of variable stars or outliers on the evaluation of results, we applied a robust filtering process. We excluded data points that deviated by more than three standard deviations from the median and iteratively repeated this process three times. In particular, for RV measurements, we also excluded the spectra of RV variable stars identified by the selection described in Section~\ref{rvv}. The gray points in Figure \ref{fig:snr_delta} represent the data identified as outliers, while the cyan points correspond to the data retained after filtering. For all parameters, the cyan points account for more than 92$\%$ of the total data, indicating that the filtered dataset is highly representative of the overall sample.

To quantify the relationship between \( \Delta P \) and S/N, we applied a fit to the filtered data using Equation\,4 from Paper\,1:

\begin{equation}
    \sigma_P = a \cdot x^b + c,
\end{equation}
where \( x \) represents the S/N. The fitting coefficients \( a \), \( b \), and \( c \), which characterize this relationship, are provided in Table \ref{tab:fitting_coefficients} for each spectral parameter, including \(T_\text{eff}\), $log\,g$, [Fe/H], $[\alpha/M]$, RV, and \(v \sin i\). The right side of Table \ref{tab:fitting_coefficients} displays the estimated internal errors at S/N values of 10, 20, and 50. As expected, internal errors decrease as S/N increases, stabilizing once the S/N exceeds 50, which indicates that higher-quality spectra yield more reliable parameter estimates in general. For example, the error in $T_{\text{eff}}$ reduces from 120 K at S/N = 10 to 27 K at S/N = 50, with similar trends observed for other parameters such as $log\,g$ and [Fe/H]. The measurement uncertainties for \( T_{\text{eff}} \) in Paper 1 are reported as 101 K at S/N = 10 and 29 K at S/N = 50, which aligns well with the uncertainties obtained in this study, providing further validation of the parameter uncertainty estimates. When the S/N reaches 20, the typical measurement uncertainties for $T_{\rm eff}$, $\log\,g$, [Fe/H], $[\alpha/{\rm M}]$, radial velocity (RV), and $v\sin i$ are approximately 59\,K, 0.09\,dex, 0.06\,dex, 0.05\,dex, 0.99\,km\,s$^{-1}$, and 3.5\,km\,s$^{-1}$, respectively. Based on this, we consider S/N~$\geq$~20 as the threshold for high-quality and reliable parameter estimates in our dataset.

\begin{table}[h!]
    \centering
    \caption{Fit coefficients for parameters and their measurement uncertainties at S/N = 10, 20, and 50.}
     \small
     \begin{tabular}{lcccccc}
        
        \toprule
        & \multicolumn{3}{c}{\textbf{Fitting coefficients}} & \multicolumn{3}{c}{\textbf{S/N}} \\
        \cmidrule(lr){2-4} \cmidrule(lr){5-7}
        & \textbf{a} & \textbf{b} & \textbf{c} & \textbf{10} & \textbf{20} & \textbf{50} \\
        \midrule
        $T_\mathrm{eff}$ (K)&1593 & -1.16 & 9.5 & 120 & 59 & 27\\
        $\log g$ (dex) &2.32 & -1.15 & 0.02 & 0.18 &0.09 &0.04 \\
        {[Fe/H]} (dex) &1.42 & -1.07 & 0.01 & 0.13 & 0.06 & 0.03 \\
        {[$\alpha$/M]} (dex) &0.56 & -0.92 & 0.01 & 0.08 & 0.05 & 0.03\\
        RV (km/s) & 139.51 & -2.08 & 0.71 & 1.9 & 0.99 & 0.75 \\
        $v \sin i$ (km/s) &41.52 & -0.02 & -36.0 & 4.0 &3.5 &2.9 \\
        \bottomrule
    \end{tabular}
    \label{tab:fitting_coefficients}
\end{table}

\subsection{External uncertainties in the parameters of LK-MRS-\uppercase\expandafter{\romannumeral1}}

To further assess the accuracy and reliability of the derived parameters, we performed a systematic analysis of their external uncertainties. Following the common practice adopted by \cite{2020Wang}, we classify stars into dwarfs ($\log g \geq 3.5$) and giants ($\log g < 3.5$) based solely on surface gravity. Applying this threshold to the matched sample from APOGEE and GALAH, we identify a total of 7,560 dwarfs and 29,014 giants. This classification enables a more precise evaluation of the external consistency of our stellar parameter estimates. The 17th data release of the Sloan Digital Sky Survey \citep{2022Abdurro'uf} and the third data release of GALAH \citep{2021Buder} provide extensive high-resolution spectroscopic data, making them ideal for cross-comparison with the parameters obtained from LK-MRS-\uppercase\expandafter{\romannumeral1}. For stellar parameters such as $T_\text{eff}$, $\log g$, [Fe/H], $v \sin i$, and $[\alpha/\mathrm{M}]$, we conduct comparisons separately for dwarfs and giants with the values provided in APOGEE and GALAH. Additionally, Gaia Data Release 3 (Gaia DR3; \citealt{2016GaiaCollaboration,2023GaiaCollaboration}) provides high-precision astrometric and radial velocity (RV) measurements, which serve as an independent benchmark for validating our RV estimates.

\begin{figure*}
    \centering
    \includegraphics[scale=0.65]{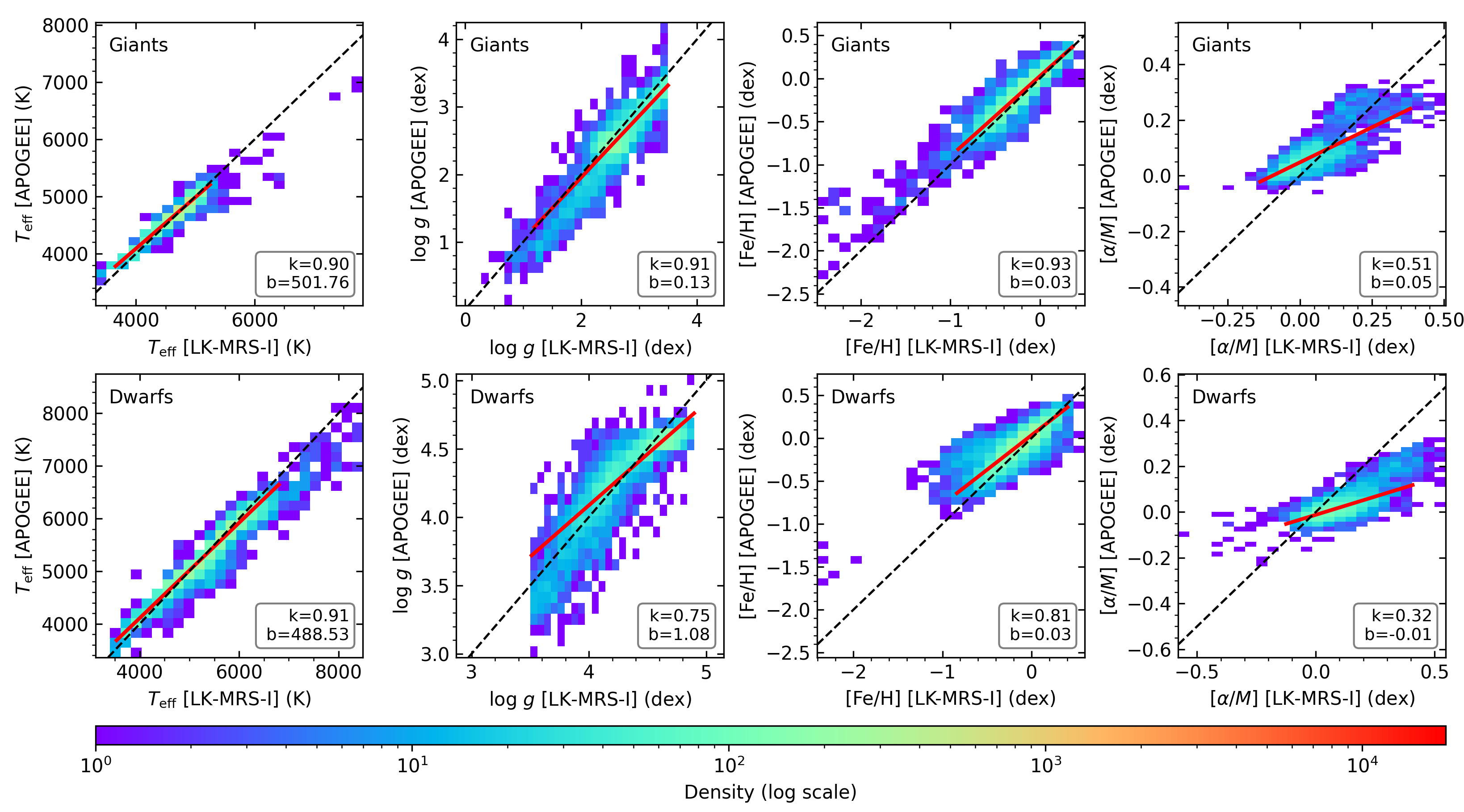}
    \caption{Comparison of stellar parameters between LK-MRS-\uppercase\expandafter{\romannumeral1} and APOGEE for giant stars (top row) and dwarf stars (bottom row). Each panel shows a 2D Gaussian density plot, where the color indicates the density of data points. The linear regression (red line) is fitted only to the high-density region that contains 80\% of all data points, in order to minimize the influence of outliers. The 1:1 reference line is shown as a black dashed line for comparison. The slope and intercept of the regression line are annotated in the lower right corner of each panel.}
    \label{fig:apogee_giant_dwarf}
\end{figure*}

\begin{figure*}
    \centering
    \includegraphics[scale=0.65]{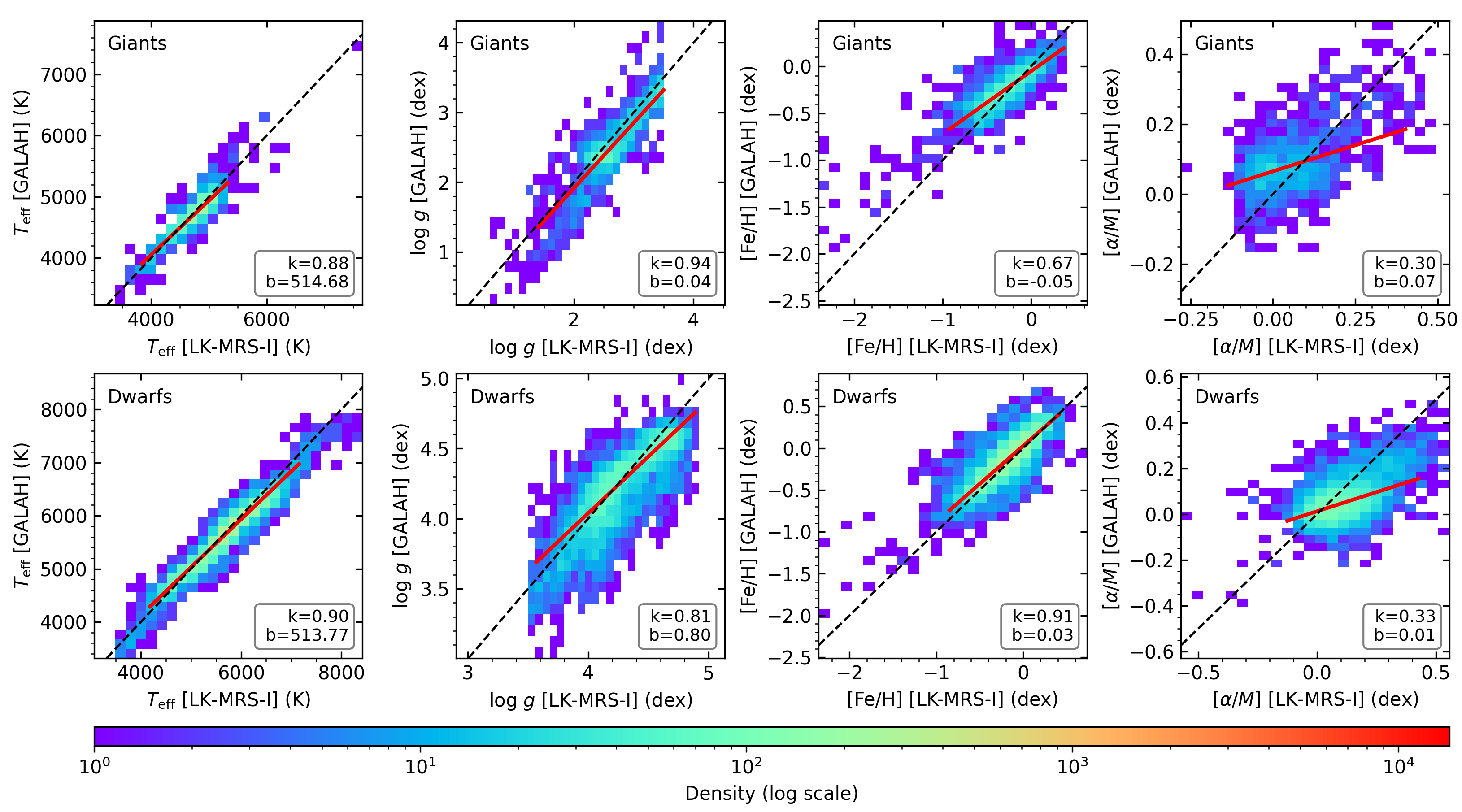}
    \caption{Same as Figure~\ref{fig:apogee_giant_dwarf}, but for the comparison between LK-MRS-\uppercase\expandafter{\romannumeral1} and GALAH. 
    }
    \label{fig:galah_giant_dwarf}
\end{figure*}

We cross-matched the targets in the processed LK-MRS-\uppercase\expandafter{\romannumeral1} dataset with those in APOGEE and GALAH using a uniform cross-matching radius of 3$\farcs$7 across all catalogs included in this study. In terms of stellar atmospheric parameters, the cross-matching with APOGEE resulted in 6,951 common sources with both \(T_\text{eff}\) and $\log\,g$, 6,859 with [Fe/H], and 4,530 with \([\alpha/\mathrm{M}]\). Similarly, the cross-match with GALAH yielded 5,394 common sources for \(T_\text{eff}\), 5,257 for both $\log\,g$ and [Fe/H], and 3,694 for \([\alpha/\mathrm{M}]\), all referring to unique stars in common with the LK-MRS-\uppercase\expandafter{\romannumeral1} dataset. Given the limited number of overlapping sources with APOGEE that include \(v \sin i\) (949 stars), we did not attempt an external uncertainty analysis for \(v \sin i\) in this work. For each atmospheric parameter we constructed a 2D Gaussian density plot in Figure~\ref{fig:apogee_giant_dwarf} and Figure~\ref{fig:galah_giant_dwarf}. Point counts in each bin were convolved with a Gaussian kernel, and the color scale encodes the logarithmic density. To minimize the impact of sparsely populated outliers, we retained only the pixels that enclose the densest 80\,\% of the cumulative distribution. A Gaussian‑weighted least‑squares fit was performed on these high‑density pixels; the resulting slope and intercept are printed in the lower‑right corner of every panel. The black dashed line marks the $1{:}1$ reference and the red line displays the final regression for the retained high‑density region.

Our statistical analysis of stellar atmospheric parameters derived from LK-MRS-\uppercase\expandafter{\romannumeral1} reveals generally good consistency with external high-resolution spectroscopic surveys, particularly APOGEE and GALAH, for key parameters including $T_{\rm eff}$, $\log g$, and [Fe/H]. As illustrated in the Figure~\ref{fig:apogee_giant_dwarf} and Figure~\ref{fig:galah_giant_dwarf}, the regression slopes for these parameters are mostly close to unity, indicating that LK-MRS-\uppercase\expandafter{\romannumeral1} provides reliable estimates across a wide range of stellar types. This consistency shows a clear dependence on stellar evolutionary stage and survey characteristics. APOGEE’s scientific focus on evolved giant stars, coupled with its near-infrared high-resolution spectroscopy ($R \sim 22{,}000$), makes its giant star parameters especially robust\citep{2019Wilson}. As a result, LK-MRS giants exhibit better agreement with APOGEE benchmarks. In contrast, the GALAH survey is designed specifically to target main-sequence stars using optical spectra (470–790 nm), which are more sensitive to effective temperature and surface gravity in dwarfs \citep{2021Buder}. Consequently, LK-MRS dwarf star parameters show improved consistency with GALAH data, exhibiting smaller systematic offsets. This multi-survey validation across different evolutionary stages and wavelength regimes demonstrates the broad applicability and reliability of LK-MRS parameter determinations.
In contrast, $[\alpha/\mathrm{M}]$ exhibits significant systematic discrepancies for both stellar types across both surveys. The regression slopes are substantially below unity, reflecting a compressed dynamic range in LK-MRS-\uppercase\expandafter{\romannumeral1} results. These deviations are likely due to the limited resolution ($R \sim 7500$) and narrow wavelength coverage of the LAMOST spectrograph, which reduces sensitivity to key $\alpha$-element absorption features such as Mg, Si, and Ca. As a result, $\alpha$-element abundances from LK-MRS-\uppercase\expandafter{\romannumeral1} should be interpreted with caution, particularly in studies requiring precise chemical abundances.

\begin{figure*}
    \centering
    \includegraphics[scale=0.45]{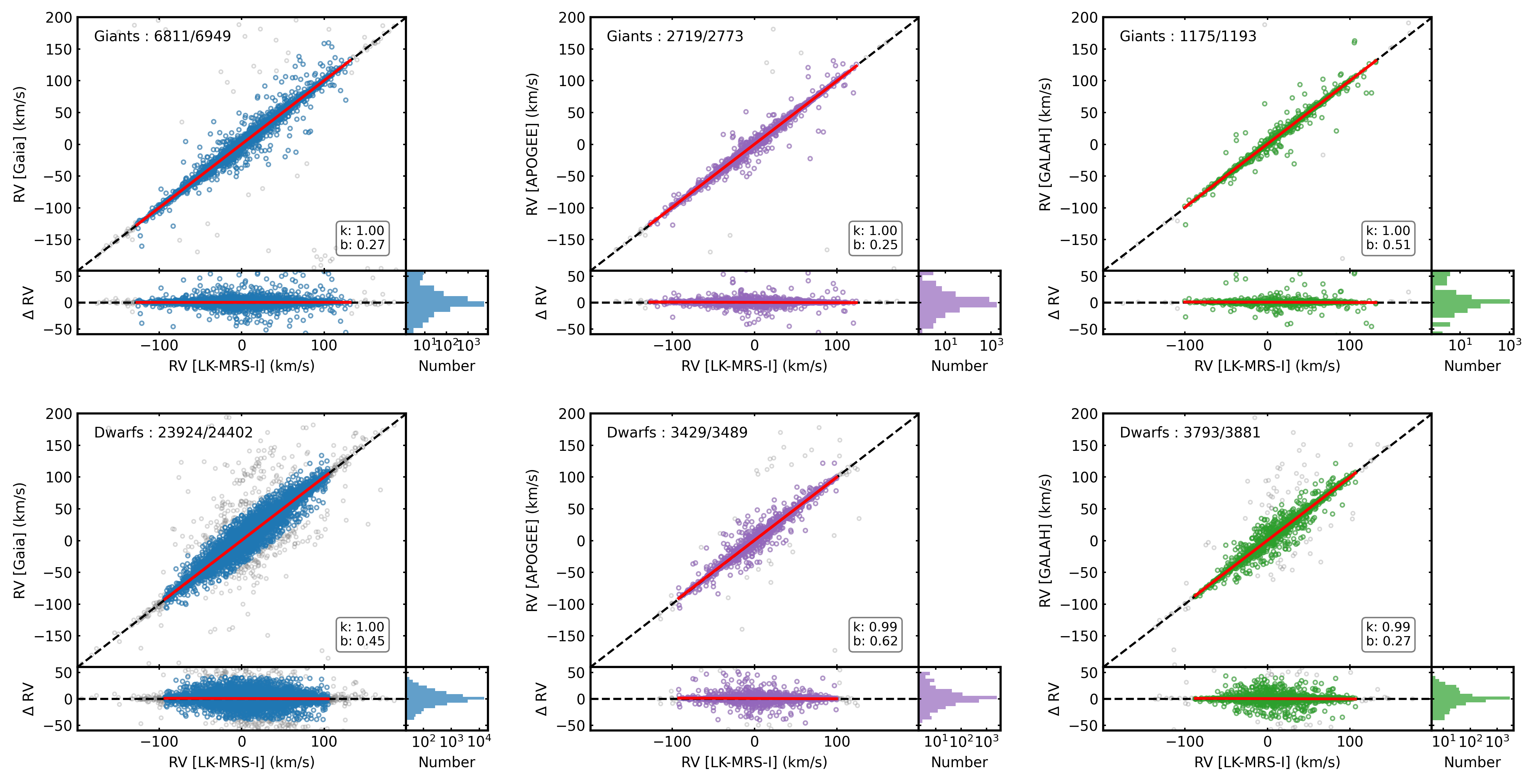}
    \caption{Comparison of RVs in the LK–MRS project for giants (top row) and dwarfs (bottom row). Each column corresponds to a survey (from left to right: Gaia, APOGEE, and GALAH). For each panel group, the top-left plot shows the RV comparison between LK-MRS-\uppercase\expandafter{\romannumeral1} and the external reference; the bottom-left plot shows the residuals ($\Delta$RV); and the bottom-right plot displays the histogram of residuals.} The black dashed line indicates the one-to-one relation, while the red solid line represents the linear regression fit to the selected data. Mean and RMS values of residuals are indicated in each plot.
    \label{fig:rv_compare}
\end{figure*}

As regards the RV, Figure~\ref{fig:rv_compare} shows a high level of consistency between LK-MRS-\uppercase\expandafter{\romannumeral1} and the reference datasets (Gaia, APOGEE, and GALAH), with regression slopes all close to unity for both giants and dwarfs. Specifically, the slopes range from 0.994 to 1.000 and the intercepts are all within $\pm$0.6\,km\,s$^{-1}$, indicating negligible systematic offsets. The residual plots further confirm that most data points lie close to zero residual, suggesting excellent agreement in RV measurements across all datasets. These results demonstrate that the RVs derived from LK-MRS-\uppercase\expandafter{\romannumeral1} are reliable and consistent with those from high-precision surveys.

\section{Peculiar Stars} \label{sec:discussion}

\subsection{Metal-poor star candidates}

\begin{figure*}
    \centering
    \begin{minipage}{1.0\textwidth} 
        \includegraphics[width=\linewidth]{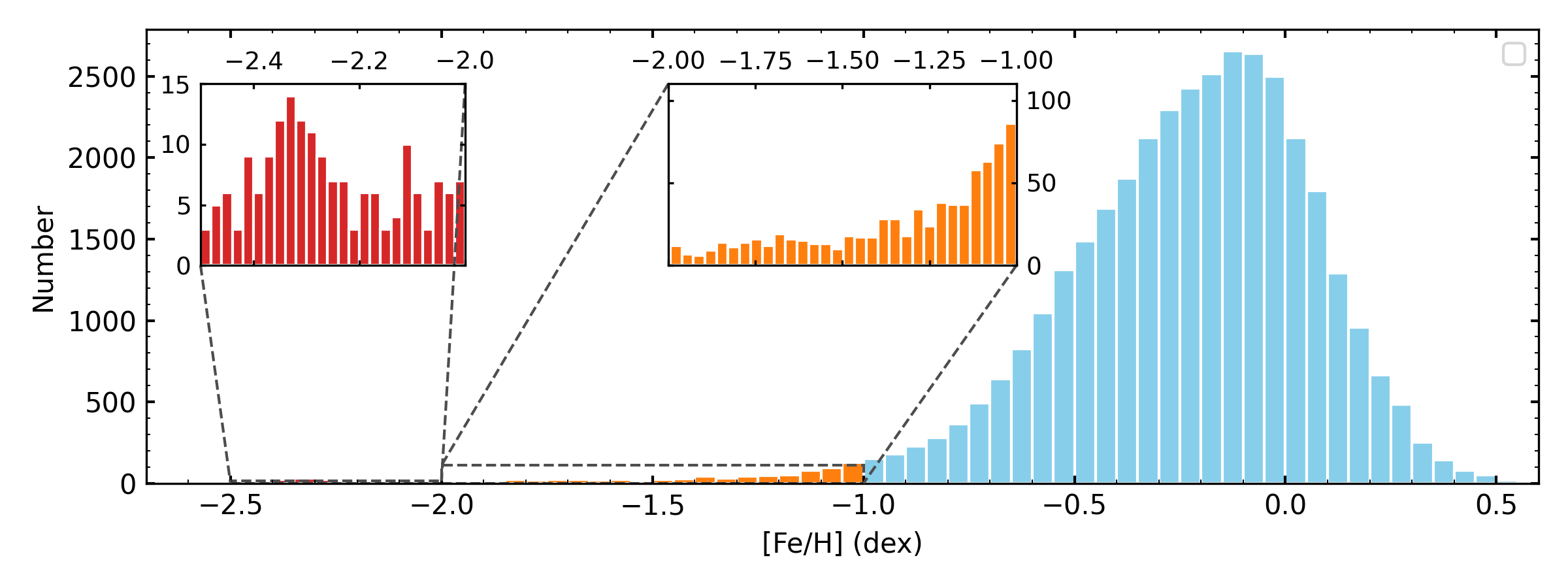}
        \caption{The histogram shows the distribution of stellar metallicities [Fe/H] for 35,614 stars measured by LK-MRS-\uppercase\expandafter{\romannumeral1}. The blue bars represent the overall metallicity distribution, while the orange and red bars highlight metal-poor (\(-2.0 \leq \text{[Fe/H]} \leq -1.0\)) and very metal-poor (\(-2.5 \leq \text{[Fe/H]} \leq -2.0\)) star candidates, respectively. The inset on the middle provides a closer view of the metal-poor range, while the inset on the left zooms in on the very metal-poor range.}
        \label{fig:feh}
    \end{minipage}
\end{figure*}

Metal-poor stars play a crucial role in probing the origin of first-generation stars in the Galaxy, tracing element production from supernovae and examining the metallicity distribution function of the Galactic halo \citep{2005Beers}. To support these investigations, acquiring a large, homogenous sample of stellar metallicities is essential. The LK-MRS-\uppercase\expandafter{\romannumeral1} measured the metallicities of 35,614 stars, covering a range from -2.5 to 0.9\,dex, as illustrated in Figure \ref{fig:feh}. The blue bars represent the overall distribution, while the orange and red bars highlight the candidates we identified as metal-poor (\(-2.0 \leq \text{[Fe/H]} \leq -1.0\)) and very metal-poor (\(-2.5 \leq \text{[Fe/H]} \leq -2.0\)) stars, numbering 746 and 174, respectively, as labeled 'MP' and 'VMP' in Table \ref{tab:parameters}. To determine which of these are newly discovered, we cross-matched our candidates with the SIMBAD database \citep{2000Wenger} to check for existing classifications. Among the 174 very metal-poor stars, 131 are newly discovered, while 43 have been previously identified in the literature. Similarly, of the 746 metal-poor stars, 678 are newly identified in this work, and 86 have existing records. The full list of MP and VMP stars is provided in Table~\ref{tab:metal_poor} in Appendix~\ref{melta_poor_tab}, and is also available online in a machine-readable format. The middle inset of Figure~\ref{fig:feh} zooms in on the MP distribution, and the left inset focuses on the VMP range, providing detailed views of these low-metallicity populations.

It is important to note that the majority of candidates for metal-poor stars or very metal-poor stars fall into the F, G, and K types. This is primarily because LASP offers limited spectral templates, only providing parameters for late A-type, F-type, G-type, and K-type stars \citep{2015Luo}. Given that our spectra have a medium resolution (approximately 7500), we can only classify these stars as candidates for metal-poor stars or very metal-poor stars. For instance, in our work, EPIC 220387907 is identified as a candidate for a very metal-poor star with an [Fe/H] value of -2.25. However, although \cite{2018Abohalima} also identifies it as a very metal-poor star, the [Fe/H] value was given as -2.75. Therefore, further confirmation is required for these candidates of metal-poor stars. Most candidates for metal-poor stars have a magnitude below 15.0 mag, making them suitable for high-resolution follow-up observations with ground-based telescopes, which would enable further confirmation.

\subsection{High-velocity star candidates}

High-velocity stars, known for their exceptional high speeds and unusual trajectories, have become key subjects in astronomical research \citep{2021Li}. The LK-MRS-\uppercase\expandafter{\romannumeral1} conducted 161,270 radial velocity measurements of 34,444 stars, covering a velocity range of -260 to 350\,km/s.

\begin{figure}[ht]
    \centering
    \begin{minipage}{0.46\textwidth} 
        \includegraphics[width=\linewidth]{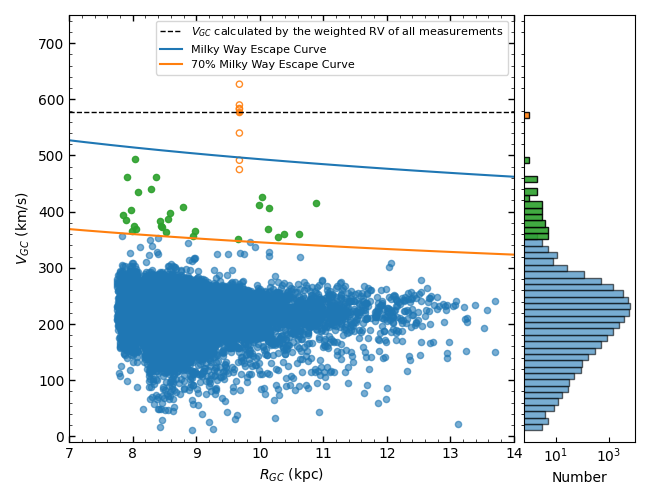}
        \caption{
        The figure illustrates the distribution of stars as a function of their galactocentric radial distance (\(R_{GC}\)) and total velocity (\(V_{GC}\)). The blue curve represents the Galactic escape velocity, while the orange curve marks 70\% of this threshold. The open circles indicate the seven individual radial velocity (RV) measurements of KIC~7881304, each converted to a corresponding \(V_{GC}\) value. The solid dot denotes the \(V_{GC}\) value derived from the weighted-mean RV of KIC~7881304, and the dashed line highlights this representative velocity. The histogram on the right presents the overall distribution of \(V_{GC}\) values.}
        \label{fig:high_velocity}
    \end{minipage}
\end{figure}

To identify high-velocity stars, we combined the proper motion and positional information provided by the Gaia DR\,3 with the radial velocity measurements from the LK-MRS. Utilizing the \texttt{Astropy} package \citep{2022Astropy}, we transformed these observational data into the three-dimensional Cartesian velocity components $(U, V, W)$. The total velocity, $V_G$, was then calculated as the square root of the sum of the squares of these velocity components. Furthermore, we employed the \texttt{Galpy} package \citep{2015Bovy}, to model the Galactic escape velocity curve, which is represented as a blue line in Figure \ref{fig:high_velocity}. The orange line in Figure \ref{fig:high_velocity}, representing 70\% of the escape velocity, acts as a boundary for distinguishing different velocity regimes of high-velocity stars \citep{2024Liao}.

Comparing the $V_G$ velocities of 34,444 stars from the LK-MRS-\uppercase\expandafter{\romannumeral1} with the Galactic escape velocity, we identified one star whose velocity exceeds this threshold. This star, known as KIC 7881304, has been classified as a long-period variable by \cite{2020Yu}. Seven LAMOST-MRS spectra indicate a variable RV ranging from 181.8 km/s to 349.2 km/s. These seven measurements are shown as open circles in Figure~\ref{fig:high_velocity}, among which five exceed the Galactic escape velocity. Based on this, we consider KIC~7881304 a possible unbound high-velocity candidate. It is labeled as ``unbound\_HiVels'' in Table~\ref{tab:parameters}. In addition, 29 stars were found to be moving faster than 70$\%$ of the Milky Way's escape velocity. Hence, they were classified as possible candidates for bound high-velocity stars. These 29 stars are labeled as "bound\_HiVels" in Table \ref{tab:parameters}. We cross-matched these candidates with the SIMBAD database to assess whether they were previously known. Among them, 26 of the 29 bound high-velocity stars are newly discovered in this work, while the remaining 3 have been identified in previous surveys. A detailed list of these 30 high-velocity star candidates, including their RV, $V_G$ and classifications, is provided in Table~\ref{tab:high_velocity} in the Appendix~\ref{high_velocity_tab}. Compared with metal-poor stars, we found that most of these 30 high-velocity star candidates exhibit low metallicity, including 9 metal-poor stars and 18 very metal-poor stars. However, the final confirmation of these high-velocity star candidates requires further investigations, along with additional observational data and theoretical modeling to ensure a comprehensive and accurate analysis.

\subsection{RV variable star candidates} \label{rvv}

RV curves are useful in astronomical research including recognizing and studying spectroscopic multiple systems \citep{2024Castro,2024Pawar}, such as binaries, triple stars, etc., identifying modes of pulsating stars \citep{2024Hocd}, and more. Building on the significance of RV curves in stellar analysis, we focus on identifying stars with RV variations using the TD spectra from LK-MRS-\uppercase\expandafter{\romannumeral1}. To ensure a robust identification of the candidates, we applied a selection criterion requiring the standard deviation of RV measurements to exceed three times the mean RV uncertainty. Using this threshold, 
we identified a sample of 2333 candidates exhibiting significant RV variability out of 34,468 stars, corresponding to $\sim$ 6.8\% of the total population. These stars are labeled as "RVV" in Table~\ref{tab:parameters}, and a complete record of their individual radial velocity measurements is provided in the Appendix~\ref{rvv_tab} (Table~\ref{tab:rv_variable}). Additionally, we indicate whether each star is classified as a periodic variable in our later analysis.

\begin{figure*}[ht]
    \centering
    \begin{minipage}{1.0\textwidth} 
        \includegraphics[width=\linewidth]{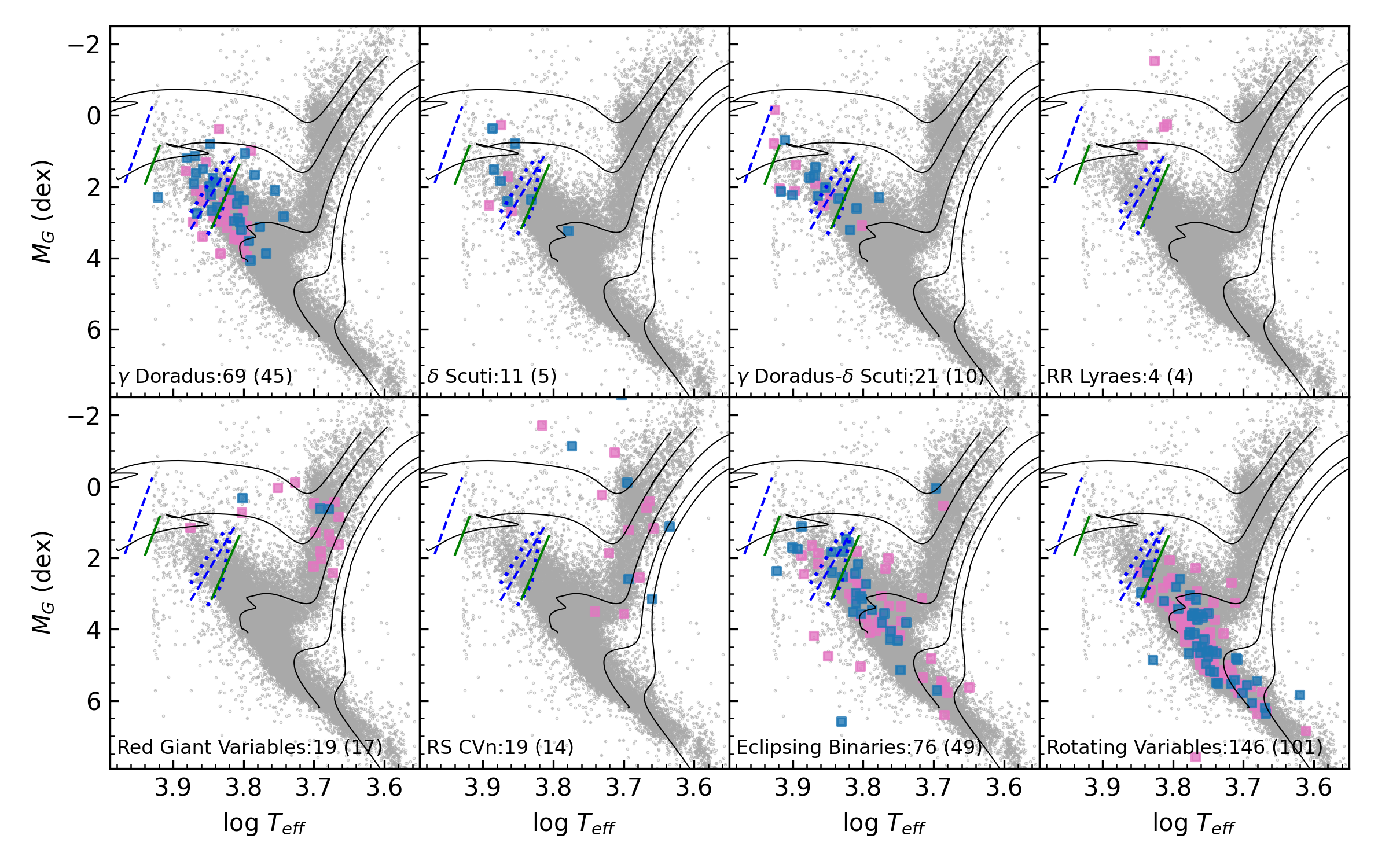}
        \caption{Effective temperature versus absolute magnitude diagram ($T_{\text{eff}}$ vs. $M_G$) for variable stars identified in the LK-MRS-\uppercase\expandafter{\romannumeral1} survey. The observational $\delta$~Sct IS is shown as a blue dashed line, the theoretical $\delta$~Sct IS as a green band, and the $\gamma$~Dor IS as a blue dotted line. Black curves represent MIST evolutionary tracks for stellar masses from 0.4 to 2.8\,M$_\odot$. Gray background points correspond to all LK-MRS-\uppercase\expandafter{\romannumeral1} targets. Variable stars classified by \textit{Kepler}/K2 and \textit{TESS}  are marked with pink and light blue squares, respectively, with their types annotated in the figure. The total number of classified variables is also indicated in each panel, with the number contributed by \textit{Kepler}/K2 shown in parentheses.}
        \label{fig:variable_star}
    \end{minipage}
\end{figure*}

Among these candidates, some stars also have photometric data from the \textit{Kepler}, \textit{K2} missions or \textit{TESS} \citep{2014Ricker}
, facilitating further classification. By cross-matching the RV variability candidates with the \textit{Kepler}, \textit{K2} and \textit{TESS} photometric database \citep{2018tess,2021tess,2021tesssc}, we found that 1088 out of the 2333 variable star candidates have \textit{Kepler}/\textit{K2} photometric data, 1709 out of the 2333 variable star candidates have \textit{TESS} photometric data. For these stars, we applied the Lomb-Scargle periodogram to extract periodic features from their light curves and calculated the S/N of the peak frequencies. Using a threshold S/N value of 5.6 \citep{2016Zong}, we successfully identified 371 periodic variables.

We further analyzed these 371 stars by examining their light curves, periodograms, and spectroscopic parameters to classify their variability types, with reference to the VSX catalog \citep{2006Watson} and the variable star catalog from \textit{Gaia} DR3 \citep{2022GaiaCollaboration} during the classification process. Among them, 194 are newly reported in this work, while 177 had been previously documented in the literature, as determined through cross-matching with the SIMBAD database. Our analysis revealed a diverse range of variable stars, including 76 eclipsing binaries, 11 $\delta$~Scuti variables, 69 $\gamma$~Doradus variables, 21 hybrid $\gamma$~Doradus and $\delta$~Scuti variables, 19 red giants with solar-like oscillations, 4 RR Lyrae variables, 19 RS Canum Venaticorum variables, and 146 rotating variable stars. The rotating variable stars include those modulated by starspots, stars exhibiting flares, and ellipsoidal variables. Since a single star may exhibit multiple phenomena, such as both starspots and flares, we collectively refer to stars whose brightness varies periodically due to rotation as rotating variable stars. Moreover, we identified six stars whose variability types remain uncertain. We note that in a few cases, our classification differs from that reported in previous surveys. These discrepancies likely arise from the use of different data sources (e.g., spectral resolution, temporal coverage). Therefore, we regard these classifications as candidate types, and recommend that more comprehensive analyses—including high-precision photometry, spectroscopy, and long-term monitoring—be conducted to confirm their true variability nature.

Using the parallaxes and apparent magnitudes provided by Gaia, along with the parameters from LK-MRS-\uppercase\expandafter{\romannumeral1}, we calculated the absolute magnitudes of these stars. Their positions on the effective temperature versus absolute magnitude diagram (T$_{\text{eff}}$ vs. $M_G$) are shown in Figure \ref{fig:variable_star}. The dashed blue belt represents the observational $\delta$ Sct instability strip (IS, \citeauthor{2019Murphy} \citeyear{2019Murphy}), while the green belt indicates the theoretical $\delta$ Sct IS \citep{2005Dupret}. The dotted blue belt within the $\delta$ Sct IS corresponds to the theoretical $\gamma$ Dor IS \citep{2005Dupret}. The black curves represent the stellar evolutionary tracks for masses of 0.4\,M$_\odot$, 0.8\,M$_\odot$, 1.2\,M$_\odot$, 2.0\,M$_\odot$, and 2.8\,M$_\odot$ from MIST \citep{2016Dotter, 2016Choi, 2011Paxton, 2013Paxton, 2015Paxton}. Although extinction corrections were applied to the magnitudes of these stars using the \texttt{dustmap} Python package \citep{2018Green}, a subset of $\gamma$ Doradus stars remains slightly offset from the theoretical instability strip. This displacement, along with the misplacement of some red giants and RS CVn stars in the parameter space, likely stems from residual uncertainties in extinction corrections or potential systematic errors in stellar parameter determinations. The classified variable stars are denoted by colored markers: pink squares represent variables identified through \textit{Kepler}/\textit{K2} light curve analysis (with their counts indicated parenthetically), while lightblue squares correspond to those classified using \textit{TESS} photometric data. These results demonstrate that the LK-MRS-\uppercase\expandafter{\romannumeral1} survey not only establishes a crucial benchmark for future asteroseismic investigations but also provides essential constraints for modeling stellar variability mechanisms across different evolutionary stages.


\section{Summary} \label{sec:summary}

As part of the second phase of LAMOST, LK-MRS-\uppercase\expandafter{\romannumeral1} has been conducting a five-year TD spectroscopic observation campaign since September 2018, focusing on 20 plates in the \textit{Kepler} and \textit{K2} fields. During this period, we obtained 3,531,785 spectra of 49,310 stars. The LASP pipeline analyzed the coadded spectra using various spectral templates, calculating \(T_\text{eff}\), \(log\,g\), and [Fe/H] for 170,637 spectra, RV for 161,270 spectra, \(v \sin i\) for 23,810 spectra, and $[\alpha/M]$ for 94,075 spectra through template matching.

We computed the weighted averages of the stellar parameters from LASP for 36,588 stars, along with their associated uncertainties. The results are summarized in Table \ref{tab:parameters}. To assess the measurement uncertainties, we calculated the differences between individual spectra and their corresponding weighted mean values, and examined their relationship with the S/N. At S/N = 10, the measurement uncertainties are as follows: 120\,K for \(T_\text{eff}\), 0.18\,dex for \(log\,g\), 0.13\,dex for [Fe/H], 0.08\,dex for $[\alpha/M]$, 1.9\,km/s for RV, and 4.0\,km/s for \(v \sin i\). As S/N increases, the parameter uncertainties decrease, stabilizing when S/N exceeds 50 in general. At S/N = 50, the uncertainties are approximately 27\,K for \(T_\text{eff}\), 0.04\,dex for \(log\,g\), 0.03\,dex for [Fe/H], 0.03\,dex for $[\alpha/M]$, 0.75\,km/s for RV, and 2.9\,km/s for \(v \sin i\). For an external uncertainty estimate, we compared our results with independent data from Gaia, APOGEE, and GALAH. The comparison showed good agreement for RV, \(T_\text{eff}\), and \(log\,g\), with regression slopes of 1.00, 0.92, and 1.03, respectively. However, systematic offsets were found for [Fe/H] and $[\alpha/M]$, with regression slopes of approximately 0.75 and 0.42, respectively.

With the metallicity information provided by LK-MRS-\uppercase\expandafter{\romannumeral1}, we identified 764 metal-poor star candidates and 174 very metal-poor star candidates. Using the Galactic escape velocity curve, we identified one high-velocity star candidate that is not gravitationally bound to the Milky Way and 29 high-velocity star candidates that are bound. Additionally, we identified 2333 stars with significant RV variations. By combining the photometric and spectroscopic data, we further classified 371 periodic variable star candidates, including 76 binaries, 11 $\delta$ Scuti stars, 69 $\gamma$ Doradus stars, 21 hybrid $\gamma$ Doradus-$\delta$ Scuti variables, 19 red giants, 4 RR Lyrae variables, 19 RS Canum Venaticorum variables, and 146 rotating variable stars. Based on the available data, we are still unable to determine the variable star type for six stars.

Since September 2023, LK-MRS has officially entered the second five-year survey. In February 2024, the LAMOST official website\footnote{https://www.lamost.org/} released an internal test version of the data resources. We are actively organizing and filtering the data collected by LK-MRS, and as astronomical observations progress, new data will be continuously released. We look forward to the scientific community making use of these datasets to drive further advancements in research.

\begin{acknowledgments}
We acknowledges the support from the National Natural Science Foundation of China (NSFC) through the grants 12090040, 12090042 and 12427804, and the science research grants from the China Manned Space Project. This work is also supported by the Central Guidance for Local Science and Technology Development Fund (No. ZYYD2025QY27). Guoshoujing Telescope (the Large Sky Area Multi-Object Fibre Spectroscopic Telescope ) is a National Major Scientific Project built by the Chinese Academy of Sciences. H.-L. Y. acknowledges the support from the Youth Innovation Promotion Association of the Chinese Academy of Sciences. Funding for the project has been provided by the National Development and Reform Commission. It is operated and managed by the National Astronomical Observatories, Chinese Academy of Sciences.

\end{acknowledgments}


\bibliography{runaway}{}

\begin{thebibliography}{}
\expandafter\ifx\csname natexlab\endcsname\relax\def\natexlab#1{#1}\fi
\providecommand{\url}[1]{\href{#1}{#1}}
\providecommand{\dodoi}[1]{doi:~\href{http://doi.org/#1}{\nolinkurl{#1}}}
\providecommand{\doeprint}[1]{\href{http://ascl.net/#1}{\nolinkurl{http://ascl.net/#1}}}
\providecommand{\doarXiv}[1]{\href{https://arxiv.org/abs/#1}{\nolinkurl{https://arxiv.org/abs/#1}}}

\bibitem[{{Abdurro'uf} {et~al.}(2022){Abdurro'uf}, {Accetta}, {Aerts}, {Silva Aguirre}, {Ahumada}, {Ajgaonkar}, {Filiz Ak}, {Alam}, {Allende Prieto}, {Almeida}, {Anders}, {Anderson}, {Andrews}, {Anguiano}, {Aquino-Ort{\'\i}z}, {Arag{\'o}n-Salamanca}, {Argudo-Fern{\'a}ndez}, {Ata}, {Aubert}, {Avila-Reese}, {Badenes}, {Barb{\'a}}, {Barger}, {Barrera-Ballesteros}, {Beaton}, {Beers}, {Belfiore}, {Bender}, {Bernardi}, {Bershady}, {Beutler}, {Bidin}, {Bird}, {Bizyaev}, {Blanc}, {Blanton}, {Boardman}, {Bolton}, {Boquien}, {Borissova}, {Bovy}, {Brandt}, {Brown}, {Brownstein}, {Brusa}, {Buchner}, {Bundy}, {Burchett}, {Bureau}, {Burgasser}, {Cabang}, {Campbell}, {Cappellari}, {Carlberg}, {Wanderley}, {Carrera}, {Cash}, {Chen}, {Chen}, {Cherinka}, {Chiappini}, {Choi}, {Chojnowski}, {Chung}, {Clerc}, {Cohen}, {Comerford}, {Comparat}, {da Costa}, {Covey}, {Crane}, {Cruz-Gonzalez}, {Culhane}, {Cunha}, {Dai}, {Damke}, {Darling}, {Davidson}, {Davies}, {Dawson}, {De Lee}, {Diamond-Stanic}, {Cano-D{\'\i}az}, {S{\'a}nchez},
  {Donor}, {Duckworth}, {Dwelly}, {Eisenstein}, {Elsworth}, {Emsellem}, {Eracleous}, {Escoffier}, {Fan}, {Farr}, {Feng}, {Fern{\'a}ndez-Trincado}, {Feuillet}, {Filipp}, {Fillingham}, {Frinchaboy}, {Fromenteau}, {Galbany}, {Garc{\'\i}a}, {Garc{\'\i}a-Hern{\'a}ndez}, {Ge}, {Geisler}, {Gelfand}, {G{\'e}ron}, {Gibson}, {Goddy}, {Godoy-Rivera}, {Grabowski}, {Green}, {Greener}, {Grier}, {Griffith}, {Guo}, {Guy}, {Hadjara}, {Harding}, {Hasselquist}, {Hayes}, {Hearty}, {Hern{\'a}ndez}, {Hill}, {Hogg}, {Holtzman}, {Horta}, {Hsieh}, {Hsu}, {Hsu}, {Huber}, {Huertas-Company}, {Hutchinson}, {Hwang}, {Ibarra-Medel}, {Chitham}, {Ilha}, {Imig}, {Jaekle}, {Jayasinghe}, {Ji}, {Johnson}, {Jones}, {J{\"o}nsson}, {Katkov}, {Khalatyan}, {Kinemuchi}, {Kisku}, {Knapen}, {Kneib}, {Kollmeier}, {Kong}, {Kounkel}, {Kreckel}, {Krishnarao}, {Lacerna}, {Lane}, {Langgin}, {Lavender}, {Law}, {Lazarz}, {Leung}, {Leung}, {Lewis}, {Li}, {Li}, {Lian}, {Liang}, {Lin}, {Lin}, {Lin}, {Lintott}, {Long}, {Longa-Pe{\~n}a}, {L{\'o}pez-Cob{\'a}}, {Lu},
  {Lundgren}, {Luo}, {Mackereth}, {de la Macorra}, {Mahadevan}, {Majewski}, {Manchado}, {Mandeville}, {Maraston}, {Margalef-Bentabol}, {Masseron}, {Masters}, {Mathur}, {McDermid}, {Mckay}, {Merloni}, {Merrifield}, {Meszaros}, {Miglio}, {Di Mille}, {Minniti}, {Minsley}, {Monachesi}, {Moon}, {Mosser}, {Mulchaey}, {Muna}, {Mu{\~n}oz}, {Myers}, {Myers}, {Nadathur}, {Nair}, {Nandra}, {Neumann}, {Newman}, {Nidever}, {Nikakhtar}, {Nitschelm}, {O'Connell}, {Garma-Oehmichen}, {Luan Souza de Oliveira}, {Olney}, {Oravetz}, {Ortigoza-Urdaneta}, {Osorio}, {Otter}, {Pace}, {Padilla}, {Pan}, {Pan}, {Parikh}, {Parker}, {Peirani}, {Pe{\~n}a Ram{\'\i}rez}, {Penny}, {Percival}, {Perez-Fournon}, {Pinsonneault}, {Poidevin}, {Poovelil}, {Price-Whelan}, {B{\'a}rbara de Andrade Queiroz}, {Raddick}, {Ray}, {Rembold}, {Riddle}, {Riffel}, {Riffel}, {Rix}, {Robin}, {Rodr{\'\i}guez-Puebla}, {Roman-Lopes}, {Rom{\'a}n-Z{\'u}{\~n}iga}, {Rose}, {Ross}, {Rossi}, {Rubin}, {Salvato}, {S{\'a}nchez}, {S{\'a}nchez-Gallego}, {Sanderson}, {Santana
  Rojas}, {Sarceno}, {Sarmiento}, {Sayres}, {Sazonova}, {Schaefer}, {Schiavon}, {Schlegel}, {Schneider}, {Schultheis}, {Schwope}, {Serenelli}, {Serna}, {Shao}, {Shapiro}, {Sharma}, {Shen}, {Shetrone}, {Shu}, {Simon}, {Skrutskie}, {Smethurst}, {Smith}, {Sobeck}, {Spoo}, {Sprague}, {Stark}, {Stassun}, {Steinmetz}, {Stello}, {Stone-Martinez}, {Storchi-Bergmann}, {Stringfellow}, {Stutz}, {Su}, {Taghizadeh-Popp}, {Talbot}, {Tayar}, {Telles}, {Teske}, {Thakar}, {Theissen}, {Tkachenko}, {Thomas}, {Tojeiro}, {Hernandez Toledo}, {Troup}, {Trump}, {Trussler}, {Turner}, {Tuttle}, {Unda-Sanzana}, {V{\'a}zquez-Mata}, {Valentini}, {Valenzuela}, {Vargas-Gonz{\'a}lez}, {Vargas-Maga{\~n}a}, {Alfaro}, {Villanova}, {Vincenzo}, {Wake}, {Warfield}, {Washington}, {Weaver}, {Weijmans}, {Weinberg}, {Weiss}, {Westfall}, {Wild}, {Wilde}, {Wilson}, {Wilson}, {Wilson}, {Wolf}, {Wood-Vasey}, {Yan}, {Zamora}, {Zasowski}, {Zhang}, {Zhao}, {Zheng}, {Zheng}, \& {Zhu}}]{2022Abdurro'uf}
{Abdurro'uf}, {Accetta}, K., {Aerts}, C., {et~al.} 2022, \apjs, 259, 35, \dodoi{10.3847/1538-4365/ac4414}

\bibitem[{{Abohalima} \& {Frebel}(2018)}]{2018Abohalima}
{Abohalima}, A., \& {Frebel}, A. 2018, \apjs, 238, 36, \dodoi{10.3847/1538-4365/aadfe9}

\bibitem[{{Astropy Collaboration} {et~al.}(2022){Astropy Collaboration}, {Price-Whelan}, {Lim}, {Earl}, {Starkman}, {Bradley}, {Shupe}, {Patil}, {Corrales}, {Brasseur}, {N{\"o}the}, {Donath}, {Tollerud}, {Morris}, {Ginsburg}, {Vaher}, {Weaver}, {Tocknell}, {Jamieson}, {van Kerkwijk}, {Robitaille}, {Merry}, {Bachetti}, {G{\"u}nther}, {Aldcroft}, {Alvarado-Montes}, {Archibald}, {B{\'o}di}, {Bapat}, {Barentsen}, {Baz{\'a}n}, {Biswas}, {Boquien}, {Burke}, {Cara}, {Cara}, {Conroy}, {Conseil}, {Craig}, {Cross}, {Cruz}, {D'Eugenio}, {Dencheva}, {Devillepoix}, {Dietrich}, {Eigenbrot}, {Erben}, {Ferreira}, {Foreman-Mackey}, {Fox}, {Freij}, {Garg}, {Geda}, {Glattly}, {Gondhalekar}, {Gordon}, {Grant}, {Greenfield}, {Groener}, {Guest}, {Gurovich}, {Handberg}, {Hart}, {Hatfield-Dodds}, {Homeier}, {Hosseinzadeh}, {Jenness}, {Jones}, {Joseph}, {Kalmbach}, {Karamehmetoglu}, {Ka{\l}uszy{\'n}ski}, {Kelley}, {Kern}, {Kerzendorf}, {Koch}, {Kulumani}, {Lee}, {Ly}, {Ma}, {MacBride}, {Maljaars}, {Muna}, {Murphy}, {Norman},
  {O'Steen}, {Oman}, {Pacifici}, {Pascual}, {Pascual-Granado}, {Patil}, {Perren}, {Pickering}, {Rastogi}, {Roulston}, {Ryan}, {Rykoff}, {Sabater}, {Sakurikar}, {Salgado}, {Sanghi}, {Saunders}, {Savchenko}, {Schwardt}, {Seifert-Eckert}, {Shih}, {Jain}, {Shukla}, {Sick}, {Simpson}, {Singanamalla}, {Singer}, {Singhal}, {Sinha}, {Sip{\H{o}}cz}, {Spitler}, {Stansby}, {Streicher}, {{\v{S}}umak}, {Swinbank}, {Taranu}, {Tewary}, {Tremblay}, {de Val-Borro}, {Van Kooten}, {Vasovi{\'c}}, {Verma}, {de Miranda Cardoso}, {Williams}, {Wilson}, {Winkel}, {Wood-Vasey}, {Xue}, {Yoachim}, {Zhang}, {Zonca}, \& {Astropy Project Contributors}}]{2022Astropy}
{Astropy Collaboration}, {Price-Whelan}, A.~M., {Lim}, P.~L., {et~al.} 2022, \apj, 935, 167, \dodoi{10.3847/1538-4357/ac7c74}

\bibitem[{{Barentsen} {et~al.}(2018){Barentsen}, {Hedges}, {Saunders}, {Cody}, {Gully-Santiago}, {Bryson}, \& {Dotson}}]{2018Barentsen}
{Barentsen}, G., {Hedges}, C., {Saunders}, N., {et~al.} 2018, arXiv e-prints, arXiv:1810.12554, \dodoi{10.48550/arXiv.1810.12554}

\bibitem[{{Beers} \& {Christlieb}(2005)}]{2005Beers}
{Beers}, T.~C., \& {Christlieb}, N. 2005, \araa, 43, 531, \dodoi{10.1146/annurev.astro.42.053102.134057}

\bibitem[{{Borucki} {et~al.}(2010){Borucki}, {Koch}, {Basri}, {Batalha}, {Brown}, {Caldwell}, {Caldwell}, {Christensen-Dalsgaard}, {Cochran}, {DeVore}, {Dunham}, {Dupree}, {Gautier}, {Geary}, {Gilliland}, {Gould}, {Howell}, {Jenkins}, {Kondo}, {Latham}, {Marcy}, {Meibom}, {Kjeldsen}, {Lissauer}, {Monet}, {Morrison}, {Sasselov}, {Tarter}, {Boss}, {Brownlee}, {Owen}, {Buzasi}, {Charbonneau}, {Doyle}, {Fortney}, {Ford}, {Holman}, {Seager}, {Steffen}, {Welsh}, {Rowe}, {Anderson}, {Buchhave}, {Ciardi}, {Walkowicz}, {Sherry}, {Horch}, {Isaacson}, {Everett}, {Fischer}, {Torres}, {Johnson}, {Endl}, {MacQueen}, {Bryson}, {Dotson}, {Haas}, {Kolodziejczak}, {Van Cleve}, {Chandrasekaran}, {Twicken}, {Quintana}, {Clarke}, {Allen}, {Li}, {Wu}, {Tenenbaum}, {Verner}, {Bruhweiler}, {Barnes}, \& {Prsa}}]{2010Borucki}
{Borucki}, W.~J., {Koch}, D., {Basri}, G., {et~al.} 2010, Science, 327, 977, \dodoi{10.1126/science.1185402}

\bibitem[{{Bovy}(2015)}]{2015Bovy}
{Bovy}, J. 2015, \apjs, 216, 29, \dodoi{10.1088/0067-0049/216/2/29}

\bibitem[{{Buder} {et~al.}(2021){Buder}, {Sharma}, {Kos}, {Amarsi}, {Nordlander}, {Lind}, {Martell}, {Asplund}, {Bland-Hawthorn}, {Casey}, {de Silva}, {D'Orazi}, {Freeman}, {Hayden}, {Lewis}, {Lin}, {Schlesinger}, {Simpson}, {Stello}, {Zucker}, {Zwitter}, {Beeson}, {Buck}, {Casagrande}, {Clark}, {{\v{C}}otar}, {da Costa}, {de Grijs}, {Feuillet}, {Horner}, {Kafle}, {Khanna}, {Kobayashi}, {Liu}, {Montet}, {Nandakumar}, {Nataf}, {Ness}, {Spina}, {Tepper-Garc{\'\i}a}, {Ting}, {Traven}, {Vogrin{\v{c}}i{\v{c}}}, {Wittenmyer}, {Wyse}, {{\v{Z}}erjal}, \& {Galah Collaboration}}]{2021Buder}
{Buder}, S., {Sharma}, S., {Kos}, J., {et~al.} 2021, \mnras, 506, 150, \dodoi{10.1093/mnras/stab1242}

\bibitem[{{Castro-Tapia} {et~al.}(2024){Castro-Tapia}, {Aguilera-G{\'o}mez}, \& {Chanam{\'e}}}]{2024Castro}
{Castro-Tapia}, M., {Aguilera-G{\'o}mez}, C., \& {Chanam{\'e}}, J. 2024, \aap, 690, A367, \dodoi{10.1051/0004-6361/202349106}

\bibitem[{{Choi} {et~al.}(2016){Choi}, {Dotter}, {Conroy}, {Cantiello}, {Paxton}, \& {Johnson}}]{2016Choi}
{Choi}, J., {Dotter}, A., {Conroy}, C., {et~al.} 2016, \apj, 823, 102, \dodoi{10.3847/0004-637X/823/2/102}

\bibitem[{{De Cat} {et~al.}(2015){De Cat}, {Fu}, {Ren}, {Yang}, {Shi}, {Luo}, {Yang}, {Wang}, {Zhang}, {Shi}, {Zhang}, {Dong}, {Catanzaro}, {Corbally}, {Frasca}, {Gray}, {Molenda-{\.Z}akowicz}, {Uytterhoeven}, {Briquet}, {Bruntt}, {Frandsen}, {Kiss}, {Kurtz}, {Marconi}, {Niemczura}, {{\O}stensen}, {Ripepi}, {Smalley}, {Southworth}, {Szab{\'o}}, {Telting}, {Karoff}, {Silva Aguirre}, {Wu}, {Hou}, {Jin}, \& {Zhou}}]{2015DeCat}
{De Cat}, P., {Fu}, J.~N., {Ren}, A.~B., {et~al.} 2015, \apjs, 220, 19, \dodoi{10.1088/0067-0049/220/1/19}

\bibitem[{{Dotter}(2016)}]{2016Dotter}
{Dotter}, A. 2016, \apjs, 222, 8, \dodoi{10.3847/0067-0049/222/1/8}

\bibitem[{{Dupret} {et~al.}(2005){Dupret}, {Grigahc{\`e}ne}, {Garrido}, {Gabriel}, \& {Scuflaire}}]{2005Dupret}
{Dupret}, M.~A., {Grigahc{\`e}ne}, A., {Garrido}, R., {Gabriel}, M., \& {Scuflaire}, R. 2005, \aap, 435, 927, \dodoi{10.1051/0004-6361:20041817}

\bibitem[{{Frasca} {et~al.}(2025){Frasca}, {Zhang}, {Alonso-Santiago}, {Fu}, {Molenda-{\.Z}akowicz}, {De Cat}, \& {Catanzaro}}]{Frasca2025}
{Frasca}, A., {Zhang}, J.~Y., {Alonso-Santiago}, J., {et~al.} 2025, \aap, 698, A7, \dodoi{10.1051/0004-6361/202553673}

\bibitem[{{Frasca} {et~al.}(2022){Frasca}, {Molenda-{\.Z}akowicz}, {Alonso-Santiago}, {Catanzaro}, {De Cat}, {Fu}, {Zong}, {Wang}, {Cang}, \& {Wang}}]{2022Frasca}
{Frasca}, A., {Molenda-{\.Z}akowicz}, J., {Alonso-Santiago}, J., {et~al.} 2022, \aap, 664, A78, \dodoi{10.1051/0004-6361/202243268}

\bibitem[{{Fu} {et~al.}(2020){Fu}, {De Cat}, {Zong}, {Frasca}, {Gray}, {Ren}, {Molenda-{\.Z}akowicz}, {Corbally}, {Catanzaro}, {Shi}, {Luo}, \& {Zhang}}]{2020Fu}
{Fu}, J.-N., {De Cat}, P., {Zong}, W., {et~al.} 2020, Research in Astronomy and Astrophysics, 20, 167, \dodoi{10.1088/1674-4527/20/10/167}

\bibitem[{{Gaia Collaboration}(2022)}]{2022GaiaCollaboration}
{Gaia Collaboration}. 2022, {VizieR Online Data Catalog: Gaia DR3 Part 4. Variability (Gaia Collaboration, 2022)}, VizieR On-line Data Catalog: I/358. Originally published in: Astron. Astrophys., in prep. (2022)

\bibitem[{{Gaia Collaboration} {et~al.}(2016){Gaia Collaboration}, {Prusti}, {de Bruijne}, {Brown}, {Vallenari}, {Babusiaux}, {Bailer-Jones}, {Bastian}, {Biermann}, {Evans}, {Eyer}, {Jansen}, {Jordi}, {Klioner}, {Lammers}, {Lindegren}, {Luri}, {Mignard}, {Milligan}, {Panem}, {Poinsignon}, {Pourbaix}, {Randich}, {Sarri}, {Sartoretti}, {Siddiqui}, {Soubiran}, {Valette}, {van Leeuwen}, {Walton}, {Aerts}, {Arenou}, {Cropper}, {Drimmel}, {H{\o}g}, {Katz}, {Lattanzi}, {O'Mullane}, {Grebel}, {Holland}, {Huc}, {Passot}, {Bramante}, {Cacciari}, {Casta{\~n}eda}, {Chaoul}, {Cheek}, {De Angeli}, {Fabricius}, {Guerra}, {Hern{\'a}ndez}, {Jean-Antoine-Piccolo}, {Masana}, {Messineo}, {Mowlavi}, {Nienartowicz}, {Ord{\'o}{\~n}ez-Blanco}, {Panuzzo}, {Portell}, {Richards}, {Riello}, {Seabroke}, {Tanga}, {Th{\'e}venin}, {Torra}, {Els}, {Gracia-Abril}, {Comoretto}, {Garcia-Reinaldos}, {Lock}, {Mercier}, {Altmann}, {Andrae}, {Astraatmadja}, {Bellas-Velidis}, {Benson}, {Berthier}, {Blomme}, {Busso}, {Carry}, {Cellino}, {Clementini},
  {Cowell}, {Creevey}, {Cuypers}, {Davidson}, {De Ridder}, {de Torres}, {Delchambre}, {Dell'Oro}, {Ducourant}, {Fr{\'e}mat}, {Garc{\'\i}a-Torres}, {Gosset}, {Halbwachs}, {Hambly}, {Harrison}, {Hauser}, {Hestroffer}, {Hodgkin}, {Huckle}, {Hutton}, {Jasniewicz}, {Jordan}, {Kontizas}, {Korn}, {Lanzafame}, {Manteiga}, {Moitinho}, {Muinonen}, {Osinde}, {Pancino}, {Pauwels}, {Petit}, {Recio-Blanco}, {Robin}, {Sarro}, {Siopis}, {Smith}, {Smith}, {Sozzetti}, {Thuillot}, {van Reeven}, {Viala}, {Abbas}, {Abreu Aramburu}, {Accart}, {Aguado}, {Allan}, {Allasia}, {Altavilla}, {{\'A}lvarez}, {Alves}, {Anderson}, {Andrei}, {Anglada Varela}, {Antiche}, {Antoja}, {Ant{\'o}n}, {Arcay}, {Atzei}, {Ayache}, {Bach}, {Baker}, {Balaguer-N{\'u}{\~n}ez}, {Barache}, {Barata}, {Barbier}, {Barblan}, {Baroni}, {Barrado y Navascu{\'e}s}, {Barros}, {Barstow}, {Becciani}, {Bellazzini}, {Bellei}, {Bello Garc{\'\i}a}, {Belokurov}, {Bendjoya}, {Berihuete}, {Bianchi}, {Bienaym{\'e}}, {Billebaud}, {Blagorodnova}, {Blanco-Cuaresma}, {Boch},
  {Bombrun}, {Borrachero}, {Bouquillon}, {Bourda}, {Bouy}, {Bragaglia}, {Breddels}, {Brouillet}, {Br{\"u}semeister}, {Bucciarelli}, {Budnik}, {Burgess}, {Burgon}, {Burlacu}, {Busonero}, {Buzzi}, {Caffau}, {Cambras}, {Campbell}, {Cancelliere}, {Cantat-Gaudin}, {Carlucci}, {Carrasco}, {Castellani}, {Charlot}, {Charnas}, {Charvet}, {Chassat}, {Chiavassa}, {Clotet}, {Cocozza}, {Collins}, {Collins}, {Costigan}, {Crifo}, {Cross}, {Crosta}, {Crowley}, {Dafonte}, {Damerdji}, {Dapergolas}, {David}, {David}, {De Cat}, {de Felice}, {de Laverny}, {De Luise}, {De March}, {de Martino}, {de Souza}, {Debosscher}, {del Pozo}, {Delbo}, {Delgado}, {Delgado}, {di Marco}, {Di Matteo}, {Diakite}, {Distefano}, {Dolding}, {Dos Anjos}, {Drazinos}, {Dur{\'a}n}, {Dzigan}, {Ecale}, {Edvardsson}, {Enke}, {Erdmann}, {Escolar}, {Espina}, {Evans}, {Eynard Bontemps}, {Fabre}, {Fabrizio}, {Faigler}, {Falc{\~a}o}, {Farr{\`a}s Casas}, {Faye}, {Federici}, {Fedorets}, {Fern{\'a}ndez-Hern{\'a}ndez}, {Fernique}, {Fienga}, {Figueras}, {Filippi},
  {Findeisen}, {Fonti}, {Fouesneau}, {Fraile}, {Fraser}, {Fuchs}, {Furnell}, {Gai}, {Galleti}, {Galluccio}, {Garabato}, {Garc{\'\i}a-Sedano}, {Gar{\'e}}, {Garofalo}, {Garralda}, {Gavras}, {Gerssen}, {Geyer}, {Gilmore}, {Girona}, {Giuffrida}, {Gomes}, {Gonz{\'a}lez-Marcos}, {Gonz{\'a}lez-N{\'u}{\~n}ez}, {Gonz{\'a}lez-Vidal}, {Granvik}, {Guerrier}, {Guillout}, {Guiraud}, {G{\'u}rpide}, {Guti{\'e}rrez-S{\'a}nchez}, {Guy}, {Haigron}, {Hatzidimitriou}, {Haywood}, {Heiter}, {Helmi}, {Hobbs}, {Hofmann}, {Holl}, {Holland}, {Hunt}, {Hypki}, {Icardi}, {Irwin}, {Jevardat de Fombelle}, {Jofr{\'e}}, {Jonker}, {Jorissen}, {Julbe}, {Karampelas}, {Kochoska}, {Kohley}, {Kolenberg}, {Kontizas}, {Koposov}, {Kordopatis}, {Koubsky}, {Kowalczyk}, {Krone-Martins}, {Kudryashova}, {Kull}, {Bachchan}, {Lacoste-Seris}, {Lanza}, {Lavigne}, {Le Poncin-Lafitte}, {Lebreton}, {Lebzelter}, {Leccia}, {Leclerc}, {Lecoeur-Taibi}, {Lemaitre}, {Lenhardt}, {Leroux}, {Liao}, {Licata}, {Lindstr{\o}m}, {Lister}, {Livanou}, {Lobel}, {L{\"o}ffler},
  {L{\'o}pez}, {Lopez-Lozano}, {Lorenz}, {Loureiro}, {MacDonald}, {Magalh{\~a}es Fernandes}, {Managau}, {Mann}, {Mantelet}, {Marchal}, {Marchant}, {Marconi}, {Marie}, {Marinoni}, {Marrese}, {Marschalk{\'o}}, {Marshall}, {Mart{\'\i}n-Fleitas}, {Martino}, {Mary}, {Matijevi{\v{c}}}, {Mazeh}, {McMillan}, {Messina}, {Mestre}, {Michalik}, {Millar}, {Miranda}, {Molina}, {Molinaro}, {Molinaro}, {Moln{\'a}r}, {Moniez}, {Montegriffo}, {Monteiro}, {Mor}, {Mora}, {Morbidelli}, {Morel}, {Morgenthaler}, {Morley}, {Morris}, {Mulone}, {Muraveva}, {Musella}, {Narbonne}, {Nelemans}, {Nicastro}, {Noval}, {Ord{\'e}novic}, {Ordieres-Mer{\'e}}, {Osborne}, {Pagani}, {Pagano}, {Pailler}, {Palacin}, {Palaversa}, {Parsons}, {Paulsen}, {Pecoraro}, {Pedrosa}, {Pentik{\"a}inen}, {Pereira}, {Pichon}, {Piersimoni}, {Pineau}, {Plachy}, {Plum}, {Poujoulet}, {Pr{\v{s}}a}, {Pulone}, {Ragaini}, {Rago}, {Rambaux}, {Ramos-Lerate}, {Ranalli}, {Rauw}, {Read}, {Regibo}, {Renk}, {Reyl{\'e}}, {Ribeiro}, {Rimoldini}, {Ripepi}, {Riva}, {Rixon},
  {Roelens}, {Romero-G{\'o}mez}, {Rowell}, {Royer}, {Rudolph}, {Ruiz-Dern}, {Sadowski}, {Sagrist{\`a} Sell{\'e}s}, {Sahlmann}, {Salgado}, {Salguero}, {Sarasso}, {Savietto}, {Schnorhk}, {Schultheis}, {Sciacca}, {Segol}, {Segovia}, {Segransan}, {Serpell}, {Shih}, {Smareglia}, {Smart}, {Smith}, {Solano}, {Solitro}, {Sordo}, {Soria Nieto}, {Souchay}, {Spagna}, {Spoto}, {Stampa}, {Steele}, {Steidelm{\"u}ller}, {Stephenson}, {Stoev}, {Suess}, {S{\"u}veges}, {Surdej}, {Szabados}, {Szegedi-Elek}, {Tapiador}, {Taris}, {Tauran}, {Taylor}, {Teixeira}, {Terrett}, {Tingley}, {Trager}, {Turon}, {Ulla}, {Utrilla}, {Valentini}, {van Elteren}, {Van Hemelryck}, {van Leeuwen}, {Varadi}, {Vecchiato}, {Veljanoski}, {Via}, {Vicente}, {Vogt}, {Voss}, {Votruba}, {Voutsinas}, {Walmsley}, {Weiler}, {Weingrill}, {Werner}, {Wevers}, {Whitehead}, {Wyrzykowski}, {Yoldas}, {{\v{Z}}erjal}, {Zucker}, {Zurbach}, {Zwitter}, {Alecu}, {Allen}, {Allende Prieto}, {Amorim}, {Anglada-Escud{\'e}}, {Arsenijevic}, {Azaz}, {Balm}, {Beck}, {Bernstein},
  {Bigot}, {Bijaoui}, {Blasco}, {Bonfigli}, {Bono}, {Boudreault}, {Bressan}, {Brown}, {Brunet}, {Bunclark}, {Buonanno}, {Butkevich}, {Carret}, {Carrion}, {Chemin}, {Ch{\'e}reau}, {Corcione}, {Darmigny}, {de Boer}, {de Teodoro}, {de Zeeuw}, {Delle Luche}, {Domingues}, {Dubath}, {Fodor}, {Fr{\'e}zouls}, {Fries}, {Fustes}, {Fyfe}, {Gallardo}, {Gallegos}, {Gardiol}, {Gebran}, {Gomboc}, {G{\'o}mez}, {Grux}, {Gueguen}, {Heyrovsky}, {Hoar}, {Iannicola}, {Isasi Parache}, {Janotto}, {Joliet}, {Jonckheere}, {Keil}, {Kim}, {Klagyivik}, {Klar}, {Knude}, {Kochukhov}, {Kolka}, {Kos}, {Kutka}, {Lainey}, {LeBouquin}, {Liu}, {Loreggia}, {Makarov}, {Marseille}, {Martayan}, {Martinez-Rubi}, {Massart}, {Meynadier}, {Mignot}, {Munari}, {Nguyen}, {Nordlander}, {Ocvirk}, {O'Flaherty}, {Olias Sanz}, {Ortiz}, {Osorio}, {Oszkiewicz}, {Ouzounis}, {Palmer}, {Park}, {Pasquato}, {Peltzer}, {Peralta}, {P{\'e}turaud}, {Pieniluoma}, {Pigozzi}, {Poels}, {Prat}, {Prod'homme}, {Raison}, {Rebordao}, {Risquez}, {Rocca-Volmerange}, {Rosen},
  {Ruiz-Fuertes}, {Russo}, {Sembay}, {Serraller Vizcaino}, {Short}, {Siebert}, {Silva}, {Sinachopoulos}, {Slezak}, {Soffel}, {Sosnowska}, {Strai{\v{z}}ys}, {ter Linden}, {Terrell}, {Theil}, {Tiede}, {Troisi}, {Tsalmantza}, {Tur}, {Vaccari}, {Vachier}, {Valles}, {Van Hamme}, {Veltz}, {Virtanen}, {Wallut}, {Wichmann}, {Wilkinson}, {Ziaeepour}, \& {Zschocke}}]{2016GaiaCollaboration}
{Gaia Collaboration}, {Prusti}, T., {de Bruijne}, J.~H.~J., {et~al.} 2016, \aap, 595, A1, \dodoi{10.1051/0004-6361/201629272}

\bibitem[{{Gaia Collaboration} {et~al.}(2023){Gaia Collaboration}, {Vallenari}, {Brown}, {Prusti}, {de Bruijne}, {Arenou}, {Babusiaux}, {Biermann}, {Creevey}, {Ducourant}, {Evans}, {Eyer}, {Guerra}, {Hutton}, {Jordi}, {Klioner}, {Lammers}, {Lindegren}, {Luri}, {Mignard}, {Panem}, {Pourbaix}, {Randich}, {Sartoretti}, {Soubiran}, {Tanga}, {Walton}, {Bailer-Jones}, {Bastian}, {Drimmel}, {Jansen}, {Katz}, {Lattanzi}, {van Leeuwen}, {Bakker}, {Cacciari}, {Casta{\~n}eda}, {De Angeli}, {Fabricius}, {Fouesneau}, {Fr{\'e}mat}, {Galluccio}, {Guerrier}, {Heiter}, {Masana}, {Messineo}, {Mowlavi}, {Nicolas}, {Nienartowicz}, {Pailler}, {Panuzzo}, {Riclet}, {Roux}, {Seabroke}, {Sordo}, {Th{\'e}venin}, {Gracia-Abril}, {Portell}, {Teyssier}, {Altmann}, {Andrae}, {Audard}, {Bellas-Velidis}, {Benson}, {Berthier}, {Blomme}, {Burgess}, {Busonero}, {Busso}, {C{\'a}novas}, {Carry}, {Cellino}, {Cheek}, {Clementini}, {Damerdji}, {Davidson}, {de Teodoro}, {Nu{\~n}ez Campos}, {Delchambre}, {Dell'Oro}, {Esquej},
  {Fern{\'a}ndez-Hern{\'a}ndez}, {Fraile}, {Garabato}, {Garc{\'\i}a-Lario}, {Gosset}, {Haigron}, {Halbwachs}, {Hambly}, {Harrison}, {Hern{\'a}ndez}, {Hestroffer}, {Hodgkin}, {Holl}, {Jan{\ss}en}, {Jevardat de Fombelle}, {Jordan}, {Krone-Martins}, {Lanzafame}, {L{\"o}ffler}, {Marchal}, {Marrese}, {Moitinho}, {Muinonen}, {Osborne}, {Pancino}, {Pauwels}, {Recio-Blanco}, {Reyl{\'e}}, {Riello}, {Rimoldini}, {Roegiers}, {Rybizki}, {Sarro}, {Siopis}, {Smith}, {Sozzetti}, {Utrilla}, {van Leeuwen}, {Abbas}, {{\'A}brah{\'a}m}, {Abreu Aramburu}, {Aerts}, {Aguado}, {Ajaj}, {Aldea-Montero}, {Altavilla}, {{\'A}lvarez}, {Alves}, {Anders}, {Anderson}, {Anglada Varela}, {Antoja}, {Baines}, {Baker}, {Balaguer-N{\'u}{\~n}ez}, {Balbinot}, {Balog}, {Barache}, {Barbato}, {Barros}, {Barstow}, {Bartolom{\'e}}, {Bassilana}, {Bauchet}, {Becciani}, {Bellazzini}, {Berihuete}, {Bernet}, {Bertone}, {Bianchi}, {Binnenfeld}, {Blanco-Cuaresma}, {Blazere}, {Boch}, {Bombrun}, {Bossini}, {Bouquillon}, {Bragaglia}, {Bramante}, {Breedt},
  {Bressan}, {Brouillet}, {Brugaletta}, {Bucciarelli}, {Burlacu}, {Butkevich}, {Buzzi}, {Caffau}, {Cancelliere}, {Cantat-Gaudin}, {Carballo}, {Carlucci}, {Carnerero}, {Carrasco}, {Casamiquela}, {Castellani}, {Castro-Ginard}, {Chaoul}, {Charlot}, {Chemin}, {Chiaramida}, {Chiavassa}, {Chornay}, {Comoretto}, {Contursi}, {Cooper}, {Cornez}, {Cowell}, {Crifo}, {Cropper}, {Crosta}, {Crowley}, {Dafonte}, {Dapergolas}, {David}, {David}, {de Laverny}, {De Luise}, {De March}, {De Ridder}, {de Souza}, {de Torres}, {del Peloso}, {del Pozo}, {Delbo}, {Delgado}, {Delisle}, {Demouchy}, {Dharmawardena}, {Di Matteo}, {Diakite}, {Diener}, {Distefano}, {Dolding}, {Edvardsson}, {Enke}, {Fabre}, {Fabrizio}, {Faigler}, {Fedorets}, {Fernique}, {Fienga}, {Figueras}, {Fournier}, {Fouron}, {Fragkoudi}, {Gai}, {Garcia-Gutierrez}, {Garcia-Reinaldos}, {Garc{\'\i}a-Torres}, {Garofalo}, {Gavel}, {Gavras}, {Gerlach}, {Geyer}, {Giacobbe}, {Gilmore}, {Girona}, {Giuffrida}, {Gomel}, {Gomez}, {Gonz{\'a}lez-N{\'u}{\~n}ez},
  {Gonz{\'a}lez-Santamar{\'\i}a}, {Gonz{\'a}lez-Vidal}, {Granvik}, {Guillout}, {Guiraud}, {Guti{\'e}rrez-S{\'a}nchez}, {Guy}, {Hatzidimitriou}, {Hauser}, {Haywood}, {Helmer}, {Helmi}, {Sarmiento}, {Hidalgo}, {Hilger}, {H{\l}adczuk}, {Hobbs}, {Holland}, {Huckle}, {Jardine}, {Jasniewicz}, {Jean-Antoine Piccolo}, {Jim{\'e}nez-Arranz}, {Jorissen}, {Juaristi Campillo}, {Julbe}, {Karbevska}, {Kervella}, {Khanna}, {Kontizas}, {Kordopatis}, {Korn}, {K{\'o}sp{\'a}l}, {Kostrzewa-Rutkowska}, {Kruszy{\'n}ska}, {Kun}, {Laizeau}, {Lambert}, {Lanza}, {Lasne}, {Le Campion}, {Lebreton}, {Lebzelter}, {Leccia}, {Leclerc}, {Lecoeur-Taibi}, {Liao}, {Licata}, {Lindstr{\o}m}, {Lister}, {Livanou}, {Lobel}, {Lorca}, {Loup}, {Madrero Pardo}, {Magdaleno Romeo}, {Managau}, {Mann}, {Manteiga}, {Marchant}, {Marconi}, {Marcos}, {Marcos Santos}, {Mar{\'\i}n Pina}, {Marinoni}, {Marocco}, {Marshall}, {Martin Polo}, {Mart{\'\i}n-Fleitas}, {Marton}, {Mary}, {Masip}, {Massari}, {Mastrobuono-Battisti}, {Mazeh}, {McMillan}, {Messina}, {Michalik},
  {Millar}, {Mints}, {Molina}, {Molinaro}, {Moln{\'a}r}, {Monari}, {Mongui{\'o}}, {Montegriffo}, {Montero}, {Mor}, {Mora}, {Morbidelli}, {Morel}, {Morris}, {Muraveva}, {Murphy}, {Musella}, {Nagy}, {Noval}, {Oca{\~n}a}, {Ogden}, {Ordenovic}, {Osinde}, {Pagani}, {Pagano}, {Palaversa}, {Palicio}, {Pallas-Quintela}, {Panahi}, {Payne-Wardenaar}, {Pe{\~n}alosa Esteller}, {Penttil{\"a}}, {Pichon}, {Piersimoni}, {Pineau}, {Plachy}, {Plum}, {Poggio}, {Pr{\v{s}}a}, {Pulone}, {Racero}, {Ragaini}, {Rainer}, {Raiteri}, {Rambaux}, {Ramos}, {Ramos-Lerate}, {Re Fiorentin}, {Regibo}, {Richards}, {Rios Diaz}, {Ripepi}, {Riva}, {Rix}, {Rixon}, {Robichon}, {Robin}, {Robin}, {Roelens}, {Rogues}, {Rohrbasser}, {Romero-G{\'o}mez}, {Rowell}, {Royer}, {Ruz Mieres}, {Rybicki}, {Sadowski}, {S{\'a}ez N{\'u}{\~n}ez}, {Sagrist{\`a} Sell{\'e}s}, {Sahlmann}, {Salguero}, {Samaras}, {Sanchez Gimenez}, {Sanna}, {Santove{\~n}a}, {Sarasso}, {Schultheis}, {Sciacca}, {Segol}, {Segovia}, {S{\'e}gransan}, {Semeux}, {Shahaf}, {Siddiqui}, {Siebert},
  {Siltala}, {Silvelo}, {Slezak}, {Slezak}, {Smart}, {Snaith}, {Solano}, {Solitro}, {Souami}, {Souchay}, {Spagna}, {Spina}, {Spoto}, {Steele}, {Steidelm{\"u}ller}, {Stephenson}, {S{\"u}veges}, {Surdej}, {Szabados}, {Szegedi-Elek}, {Taris}, {Taylor}, {Teixeira}, {Tolomei}, {Tonello}, {Torra}, {Torra}, {Torralba Elipe}, {Trabucchi}, {Tsounis}, {Turon}, {Ulla}, {Unger}, {Vaillant}, {van Dillen}, {van Reeven}, {Vanel}, {Vecchiato}, {Viala}, {Vicente}, {Voutsinas}, {Weiler}, {Wevers}, {Wyrzykowski}, {Yoldas}, {Yvard}, {Zhao}, {Zorec}, {Zucker}, \& {Zwitter}}]{2023GaiaCollaboration}
{Gaia Collaboration}, {Vallenari}, A., {Brown}, A.~G.~A., {et~al.} 2023, \aap, 674, A1, \dodoi{10.1051/0004-6361/202243940}

\bibitem[{{Green}(2018)}]{2018Green}
{Green}, G.~M. 2018, The Journal of Open Source Software, 3, 695, \dodoi{10.21105/joss.00695}

\bibitem[{{Han} {et~al.}(2023){Han}, {Wang}, {Bai}, {Yang}, {Fang}, \& {Liu}}]{2023Han}
{Han}, H., {Wang}, S., {Bai}, Y., {et~al.} 2023, \apjs, 264, 12, \dodoi{10.3847/1538-4365/ac9eac}

\bibitem[{{Hocd{\'e}} {et~al.}(2024){Hocd{\'e}}, {Moskalik}, {Gorynya}, {Smolec}, {Rathour}, \& {Zi{\'o}{\l}kowska}}]{2024Hocd}
{Hocd{\'e}}, V., {Moskalik}, P., {Gorynya}, N.~A., {et~al.} 2024, \aap, 689, A224, \dodoi{10.1051/0004-6361/202347798}

\bibitem[{{Howell} {et~al.}(2014){Howell}, {Sobeck}, {Haas}, {Still}, {Barclay}, {Mullally}, {Troeltzsch}, {Aigrain}, {Bryson}, {Caldwell}, {Chaplin}, {Cochran}, {Huber}, {Marcy}, {Miglio}, {Najita}, {Smith}, {Twicken}, \& {Fortney}}]{2014Howell}
{Howell}, S.~B., {Sobeck}, C., {Haas}, M., {et~al.} 2014, \pasp, 126, 398, \dodoi{10.1086/676406}

\bibitem[{{Jin} {et~al.}(2024){Jin}, {Fu}, {Zhang}, {Zong}, {Wang}, {Ma}, {Xing}, \& {Wang}}]{2024Jin}
{Jin}, M., {Fu}, J., {Zhang}, X., {et~al.} 2024, \aj, 168, 280, \dodoi{10.3847/1538-3881/ad8913}

\bibitem[{{Li} {et~al.}(2024){Li}, {Wang}, {Han}, {Yang}, {Zheng}, {Huang}, \& {Liu}}]{2024Li}
{Li}, X., {Wang}, S., {Han}, H., {et~al.} 2024, \apj, 966, 69, \dodoi{10.3847/1538-4357/ad3038}

\bibitem[{{Li} {et~al.}(2022){Li}, {Wang}, {Zhao}, {Bai}, {Yuan}, {Zhang}, \& {Liu}}]{2022Li}
{Li}, X., {Wang}, S., {Zhao}, X., {et~al.} 2022, \apj, 938, 78, \dodoi{10.3847/1538-4357/ac8f29}

\bibitem[{{Li} {et~al.}(2021){Li}, {Luo}, {Lu}, {Zhang}, {Li}, {Wang}, {Zuo}, {Xiang}, {Ting}, {Marchetti}, {Li}, {Wang}, {Zhang}, {Hattori}, {Zhao}, {Zhang}, \& {Zhao}}]{2021Li}
{Li}, Y.-B., {Luo}, A.~L., {Lu}, Y.-J., {et~al.} 2021, \apjs, 252, 3, \dodoi{10.3847/1538-4365/abc16e}

\bibitem[{{Liao} {et~al.}(2024){Liao}, {Du}, {Deng}, {Ye}, {Li}, {Huang}, {Shi}, \& {Ma}}]{2024Liao}
{Liao}, J., {Du}, C., {Deng}, M., {et~al.} 2024, \aj, 167, 76, \dodoi{10.3847/1538-3881/ad18c4}

\bibitem[{{Liu} {et~al.}(2020){Liu}, {Fu}, {Shi}, {Wu}, {Han}, {Chen}, {Dong}, {Zhao}, {Chen}, {Zhang}, {Bai}, {Chen}, {Cui}, {Du}, {Hsia}, {Jiang}, {Hou}, {Hou}, {Li}, {Li}, {Li}, {Liu}, {Liu}, {Luo}, {Ren}, {Tian}, {Tian}, {Wang}, {Wu}, {Xie}, {Yan}, {Yang}, {Yu}, {Zhang}, {Zhang}, {Zhang}, {Zhang}, {Zhao}, {Zhong}, {Zong}, \& {Zuo}}]{2020Liu}
{Liu}, C., {Fu}, J., {Shi}, J., {et~al.} 2020, arXiv e-prints, arXiv:2005.07210, \dodoi{10.48550/arXiv.2005.07210}

\bibitem[{{Lu} {et~al.}(2025){Lu}, {Tian}, {Zhang}, {Chen}, {Li}, {Yang}, {Wang}, {Zhang}, \& {Sun}}]{2025Lu}
{Lu}, H.-P., {Tian}, H., {Zhang}, L.-Y., {et~al.} 2025, \apjl, 978, L32, \dodoi{10.3847/2041-8213/ad93cc}

\bibitem[{{Luo} {et~al.}(2015){Luo}, {Zhao}, {Zhao}, {Deng}, {Liu}, {Jing}, {Wang}, {Zhang}, {Shi}, {Cui}, {Chu}, {Li}, {Bai}, {Wu}, {Cai}, {Cao}, {Cao}, {Carlin}, {Chen}, {Chen}, {Chen}, {Chen}, {Chen}, {Chen}, {Chen}, {Christlieb}, {Chu}, {Cui}, {Dong}, {Du}, {Fan}, {Feng}, {Fu}, {Gao}, {Gong}, {Gu}, {Guo}, {Han}, {He}, {Hou}, {Hou}, {Hou}, {Hu}, {Hu}, {Hu}, {Huo}, {Jia}, {Jiang}, {Jiang}, {Jiang}, {Jin}, {Kong}, {Kong}, {Lei}, {Li}, {Li}, {Li}, {Li}, {Li}, {Li}, {Li}, {Li}, {Li}, {Li}, {Li}, {Li}, {Liang}, {Lin}, {Liu}, {Liu}, {Liu}, {Liu}, {Lu}, {Luo}, {Mao}, {Newberg}, {Ni}, {Qi}, {Qi}, {Shen}, {Shi}, {Song}, {Song}, {Su}, {Su}, {Tang}, {Tao}, {Tian}, {Wang}, {Wang}, {Wang}, {Wang}, {Wang}, {Wang}, {Wang}, {Wang}, {Wang}, {Wang}, {Wang}, {Wang}, {Wang}, {Wang}, {Wang}, {Wang}, {Wang}, {Wang}, {Wang}, {Wang}, {Wei}, {Wei}, {Wu}, {Wu}, {Wu}, {Wu}, {Xing}, {Xu}, {Xu}, {Xu}, {Yan}, {Yang}, {Yang}, {Yang}, {Yang}, {Yao}, {Yu}, {Yuan}, {Yuan}, {Yuan}, {Yuan}, {Zhai}, {Zhang}, {Zhang}, {Zhang}, {Zhang},
  {Zhang}, {Zhang}, {Zhang}, {Zhang}, {Zhao}, {Zhou}, {Zhou}, {Zhu}, {Zhu}, {Zou}, \& {Zuo}}]{2015Luo}
{Luo}, A.~L., {Zhao}, Y.-H., {Zhao}, G., {et~al.} 2015, Research in Astronomy and Astrophysics, 15, 1095, \dodoi{10.1088/1674-4527/15/8/002}

\bibitem[{{Ma} {et~al.}(2023){Ma}, {Zong}, {Fu}, {Charpinet}, {Wang}, \& {Xing}}]{2023Ma}
{Ma}, X.~Y., {Zong}, W., {Fu}, J.~N., {et~al.} 2023, \aap, 680, A11, \dodoi{10.1051/0004-6361/202347410}

\bibitem[{{Mathur} {et~al.}(2023){Mathur}, {Claytor}, {Santos}, {Garc{\'\i}a}, {Amard}, {Bugnet}, {Corsaro}, {Bonanno}, {Breton}, {Godoy-Rivera}, {Pinsonneault}, \& {van Saders}}]{2023Mathur}
{Mathur}, S., {Claytor}, Z.~R., {Santos}, {\^A}. R.~G., {et~al.} 2023, \apj, 952, 131, \dodoi{10.3847/1538-4357/acd118}

\bibitem[{{Murphy} {et~al.}(2019){Murphy}, {Hey}, {Van Reeth}, \& {Bedding}}]{2019Murphy}
{Murphy}, S.~J., {Hey}, D., {Van Reeth}, T., \& {Bedding}, T.~R. 2019, \mnras, 485, 2380, \dodoi{10.1093/mnras/stz590}

\bibitem[{{Pan} {et~al.}(2024){Pan}, {Frasca}, {Wang}, {Fu}, \& {Zhang}}]{2024Pan}
{Pan}, Y., {Frasca}, A., {Wang}, J.-X., {Fu}, J.-N., \& {Zhang}, X.-B. 2024, \aj, 168, 253, \dodoi{10.3847/1538-3881/ad84f5}

\bibitem[{{Pan} {et~al.}(2020){Pan}, {Fu}, {Zong}, {Zhang}, {Wang}, \& {Li}}]{2020Pan}
{Pan}, Y., {Fu}, J.-N., {Zong}, W., {et~al.} 2020, \apj, 905, 67, \dodoi{10.3847/1538-4357/abc250}

\bibitem[{{Pawar} {et~al.}(2024){Pawar}, {He{\l}miniak}, {Moharana}, {Pawar}, {Pyatnytskyy}, {Lala}, \& {Konacki}}]{2024Pawar}
{Pawar}, T., {He{\l}miniak}, K.~G., {Moharana}, A., {et~al.} 2024, \aap, 691, A101, \dodoi{10.1051/0004-6361/202451126}

\bibitem[{{Paxton} {et~al.}(2011){Paxton}, {Bildsten}, {Dotter}, {Herwig}, {Lesaffre}, \& {Timmes}}]{2011Paxton}
{Paxton}, B., {Bildsten}, L., {Dotter}, A., {et~al.} 2011, \apjs, 192, 3, \dodoi{10.1088/0067-0049/192/1/3}

\bibitem[{{Paxton} {et~al.}(2013){Paxton}, {Cantiello}, {Arras}, {Bildsten}, {Brown}, {Dotter}, {Mankovich}, {Montgomery}, {Stello}, {Timmes}, \& {Townsend}}]{2013Paxton}
{Paxton}, B., {Cantiello}, M., {Arras}, P., {et~al.} 2013, \apjs, 208, 4, \dodoi{10.1088/0067-0049/208/1/4}

\bibitem[{{Paxton} {et~al.}(2015){Paxton}, {Marchant}, {Schwab}, {Bauer}, {Bildsten}, {Cantiello}, {Dessart}, {Farmer}, {Hu}, {Langer}, {Townsend}, {Townsley}, \& {Timmes}}]{2015Paxton}
{Paxton}, B., {Marchant}, P., {Schwab}, J., {et~al.} 2015, \apjs, 220, 15, \dodoi{10.1088/0067-0049/220/1/15}

\bibitem[{{Pinsonneault} {et~al.}(2014){Pinsonneault}, {Elsworth}, {Epstein}, {Hekker}, {M{\'e}sz{\'a}ros}, {Chaplin}, {Johnson}, {Garc{\'\i}a}, {Holtzman}, {Mathur}, {Garc{\'\i}a P{\'e}rez}, {Silva Aguirre}, {Girardi}, {Basu}, {Shetrone}, {Stello}, {Allende Prieto}, {An}, {Beck}, {Beers}, {Bizyaev}, {Bloemen}, {Bovy}, {Cunha}, {De Ridder}, {Frinchaboy}, {Garc{\'\i}a-Hern{\'a}ndez}, {Gilliland}, {Harding}, {Hearty}, {Huber}, {Ivans}, {Kallinger}, {Majewski}, {Metcalfe}, {Miglio}, {Mosser}, {Muna}, {Nidever}, {Schneider}, {Serenelli}, {Smith}, {Tayar}, {Zamora}, \& {Zasowski}}]{2014Pinsonneault}
{Pinsonneault}, M.~H., {Elsworth}, Y., {Epstein}, C., {et~al.} 2014, \apjs, 215, 19, \dodoi{10.1088/0067-0049/215/2/19}

\bibitem[{{Pinsonneault} {et~al.}(2018){Pinsonneault}, {Elsworth}, {Tayar}, {Serenelli}, {Stello}, {Zinn}, {Mathur}, {Garc{\'\i}a}, {Johnson}, {Hekker}, {Huber}, {Kallinger}, {M{\'e}sz{\'a}ros}, {Mosser}, {Stassun}, {Girardi}, {Rodrigues}, {Silva Aguirre}, {An}, {Basu}, {Chaplin}, {Corsaro}, {Cunha}, {Garc{\'\i}a-Hern{\'a}ndez}, {Holtzman}, {J{\"o}nsson}, {Shetrone}, {Smith}, {Sobeck}, {Stringfellow}, {Zamora}, {Beers}, {Fern{\'a}ndez-Trincado}, {Frinchaboy}, {Hearty}, \& {Nitschelm}}]{2018Pinsonneault}
{Pinsonneault}, M.~H., {Elsworth}, Y.~P., {Tayar}, J., {et~al.} 2018, \apjs, 239, 32, \dodoi{10.3847/1538-4365/aaebfd}

\bibitem[{{Ricker} {et~al.}(2014){Ricker}, {Winn}, {Vanderspek}, {Latham}, {Bakos}, {Bean}, {Berta-Thompson}, {Brown}, {Buchhave}, {Butler}, {Butler}, {Chaplin}, {Charbonneau}, {Christensen-Dalsgaard}, {Clampin}, {Deming}, {Doty}, {De Lee}, {Dressing}, {Dunham}, {Endl}, {Fressin}, {Ge}, {Henning}, {Holman}, {Howard}, {Ida}, {Jenkins}, {Jernigan}, {Johnson}, {Kaltenegger}, {Kawai}, {Kjeldsen}, {Laughlin}, {Levine}, {Lin}, {Lissauer}, {MacQueen}, {Marcy}, {McCullough}, {Morton}, {Narita}, {Paegert}, {Palle}, {Pepe}, {Pepper}, {Quirrenbach}, {Rinehart}, {Sasselov}, {Sato}, {Seager}, {Sozzetti}, {Stassun}, {Sullivan}, {Szentgyorgyi}, {Torres}, {Udry}, \& {Villasenor}}]{2014Ricker}
{Ricker}, G.~R., {Winn}, J.~N., {Vanderspek}, R., {et~al.} 2014, in Society of Photo-Optical Instrumentation Engineers (SPIE) Conference Series, Vol. 9143, Space Telescopes and Instrumentation 2014: Optical, Infrared, and Millimeter Wave, ed. J.~M. {Oschmann}, Jr., M.~{Clampin}, G.~G. {Fazio}, \& H.~A. {MacEwen}, 914320, \dodoi{10.1117/12.2063489}

\bibitem[{{Rothermich} {et~al.}(2024){Rothermich}, {Faherty}, {Bardalez-Gagliuffi}, {Schneider}, {Kirkpatrick}, {Meisner}, {Burgasser}, {Kuchner}, {Allers}, {Gagn{\'e}}, {Caselden}, {Calamari}, {Popinchalk}, {Su{\'a}rez}, {Gerasimov}, {Aganze}, {Softich}, {Hsu}, {Karpoor}, {Theissen}, {Rees}, {Cecilio-Flores-Elie}, {Cushing}, {Marocco}, {Casewell}, {Bickle}, {Hamlet}, {Allen}, {Beaulieu}, {Colin}, {Gantier}, {Gramaize}, {Jalowiczor}, {Kabatnik}, {Kiwy}, {Martin}, {Pendrill}, {Pumphrey}, {Sainio}, {Schumann}, {Stevnbak}, {Sun}, {Tanner}, {Thakur}, {Thevenot}, \& {Wedracki}}]{2024Rothermich}
{Rothermich}, A., {Faherty}, J.~K., {Bardalez-Gagliuffi}, D., {et~al.} 2024, \aj, 167, 253, \dodoi{10.3847/1538-3881/ad324e}

\bibitem[{{Serenelli} {et~al.}(2017){Serenelli}, {Johnson}, {Huber}, {Pinsonneault}, {Ball}, {Tayar}, {Silva Aguirre}, {Basu}, {Troup}, {Hekker}, {Kallinger}, {Stello}, {Davies}, {Lund}, {Mathur}, {Mosser}, {Stassun}, {Chaplin}, {Elsworth}, {Garc{\'\i}a}, {Handberg}, {Holtzman}, {Hearty}, {Garc{\'\i}a-Hern{\'a}ndez}, {Gaulme}, \& {Zamora}}]{2017Serenelli}
{Serenelli}, A., {Johnson}, J., {Huber}, D., {et~al.} 2017, \apjs, 233, 23, \dodoi{10.3847/1538-4365/aa97df}

\bibitem[{{STScI}(2011)}]{2011kepler}
{STScI}. 2011, Kepler/KIC,  STScI/MAST, \dodoi{10.17909/T9059R}

\bibitem[{{STScI}(2016{\natexlab{a}})}]{2016kepler}
---. 2016{\natexlab{a}}, Kepler/EPIC,  STScI/MAST, \dodoi{10.17909/T93W28}

\bibitem[{{STScI}(2016{\natexlab{b}})}]{2016keplerlcsc}
---. 2016{\natexlab{b}}, Kepler LC+SC, Q0-Q17,  STScI/MAST, \dodoi{10.17909/T98304}

\bibitem[{{STScI}(2016{\natexlab{c}})}]{2016k2lightcurves}
---. 2016{\natexlab{c}}, K2 Light Curves (all),  STScI/MAST, \dodoi{10.17909/T9WS3R}

\bibitem[{{STScI}(2018)}]{2018tess}
---. 2018, TESS Input Catalog and Candidate Target List,  STScI/MAST, \dodoi{10.17909/FWDT-2X66}

\bibitem[{Team(2021{\natexlab{a}})}]{2021tess}
Team, M. 2021{\natexlab{a}}, TESS Light Curves - All Sectors,  STScI/MAST, \dodoi{10.17909/T9-NMC8-F686}

\bibitem[{Team(2021{\natexlab{b}})}]{2021tesssc}
---. 2021{\natexlab{b}}, TESS "Fast" Light Curves - All Sectors,  STScI/MAST, \dodoi{10.17909/T9-ST5G-3177}

\bibitem[{{Wang} {et~al.}(2021){Wang}, {Fu}, {Zong}, {Wang}, \& {Zhang}}]{2021Wang}
{Wang}, J., {Fu}, J.-N., {Zong}, W., {Wang}, J., \& {Zhang}, B. 2021, \mnras, 506, 6117, \dodoi{10.1093/mnras/stab1705}

\bibitem[{{Wang} {et~al.}(2024){Wang}, {Pan}, {Fu}, {Zong}, {Zong}, {Cang}, {Zhang}, \& {Pan}}]{2024Wang}
{Wang}, J., {Pan}, Y., {Fu}, J., {et~al.} 2024, \aap, 690, A201, \dodoi{10.1051/0004-6361/202449484}

\bibitem[{{Wang} {et~al.}(2020){Wang}, {Fu}, {Zong}, {Smith}, {De Cat}, {Shi}, {Luo}, {Zhang}, {Frasca}, {Corbally}, {Molenda-{\.Z}akowicz}, {Catanzaro}, {Gray}, {Wang}, \& {Pan}}]{2020Wang}
{Wang}, J., {Fu}, J.-N., {Zong}, W., {et~al.} 2020, \apjs, 251, 27, \dodoi{10.3847/1538-4365/abc1ed}

\bibitem[{{Wang} {et~al.}(2023){Wang}, {Luo}, {Zhang}, {Ting}, {O'Briain}, \& {Lamost Mrs Collaboration}}]{2023Wang}
{Wang}, R., {Luo}, A.~L., {Zhang}, S., {et~al.} 2023, \apjs, 266, 40, \dodoi{10.3847/1538-4365/acce36}

\bibitem[{{Wang} {et~al.}(2019){Wang}, {Luo}, {Chen}, {Bai}, {Chen}, {Chen}, {Dong}, {Du}, {Fu}, {Han}, {Hou}, {Hou}, {Hou}, {Jiang}, {Kong}, {Li}, {Liu}, {Liu}, {Qin}, {Shi}, {Tian}, {Wu}, {Wu}, {Xie}, {Zhang}, {Zhang}, {Zhao}, {Zhao}, {Zhong}, {Zong}, \& {Zuo}}]{2019Wang}
{Wang}, R., {Luo}, A.~L., {Chen}, J.~J., {et~al.} 2019, \apjs, 244, 27, \dodoi{10.3847/1538-4365/ab3cc0}

\bibitem[{{Watson} {et~al.}(2006){Watson}, {Henden}, \& {Price}}]{2006Watson}
{Watson}, C.~L., {Henden}, A.~A., \& {Price}, A. 2006, Society for Astronomical Sciences Annual Symposium, 25, 47

\bibitem[{{Wenger} {et~al.}(2000){Wenger}, {Ochsenbein}, {Egret}, {Dubois}, {Bonnarel}, {Borde}, {Genova}, {Jasniewicz}, {Lalo{\"e}}, {Lesteven}, \& {Monier}}]{2000Wenger}
{Wenger}, M., {Ochsenbein}, F., {Egret}, D., {et~al.} 2000, \aaps, 143, 9, \dodoi{10.1051/aas:2000332}

\bibitem[{{Wilson} {et~al.}(2019){Wilson}, {Hearty}, {Skrutskie}, {Majewski}, {Holtzman}, {Eisenstein}, {Gunn}, {Blank}, {Henderson}, {Smee}, {Nelson}, {Nidever}, {Arns}, {Barkhouser}, {Barr}, {Beland}, {Bershady}, {Blanton}, {Brunner}, {Burton}, {Carey}, {Carr}, {Colque}, {Crane}, {Damke}, {Davidson}, {Dean}, {Di Mille}, {Don}, {Ebelke}, {Evans}, {Fitzgerald}, {Gillespie}, {Hall}, {Harding}, {Harding}, {Hammond}, {Hancock}, {Harrison}, {Hope}, {Horne}, {Karakla}, {Lam}, {Leger}, {MacDonald}, {Maseman}, {Matsunari}, {Melton}, {Mitcheltree}, {O'Brien}, {O'Connell}, {Patten}, {Richardson}, {Rieke}, {Rieke}, {Roman-Lopes}, {Schiavon}, {Sobeck}, {Stolberg}, {Stoll}, {Tembe}, {Trujillo}, {Uomoto}, {Vernieri}, {Walker}, {Weinberg}, {Young}, {Anthony-Brumfield}, {Bizyaev}, {Breslauer}, {De Lee}, {Downey}, {Halverson}, {Huehnerhoff}, {Klaene}, {Leon}, {Long}, {Mahadevan}, {Malanushenko}, {Nguyen}, {Owen}, {S{\'a}nchez-Gallego}, {Sayres}, {Shane}, {Shectman}, {Shetrone}, {Skinner}, {Stauffer}, \& {Zhao}}]{2019Wilson}
{Wilson}, J.~C., {Hearty}, F.~R., {Skrutskie}, M.~F., {et~al.} 2019, \pasp, 131, 055001, \dodoi{10.1088/1538-3873/ab0075}

\bibitem[{{Wittenmyer} {et~al.}(2018){Wittenmyer}, {Sharma}, {Stello}, {Buder}, {Kos}, {Asplund}, {Duong}, {Lin}, {Lind}, {Ness}, {Zwitter}, {Horner}, {Clark}, {Kane}, {Huber}, {Bland-Hawthorn}, {Casey}, {De Silva}, {D'Orazi}, {Freeman}, {Martell}, {Simpson}, {Zucker}, {Anguiano}, {Casagrande}, {Esdaile}, {Hon}, {Ireland}, {Kafle}, {Khanna}, {Marshall}, {Saddon}, {Traven}, \& {Wright}}]{2018Wittenmyer}
{Wittenmyer}, R.~A., {Sharma}, S., {Stello}, D., {et~al.} 2018, \aj, 155, 84, \dodoi{10.3847/1538-3881/aaa3e4}

\bibitem[{{Worley} {et~al.}(2020){Worley}, {Jofr{\'e}}, {Rendle}, {Miglio}, {Magrini}, {Feuillet}, {Gavel}, {Smiljanic}, {Lind}, {Korn}, {Gilmore}, {Randich}, {Hourihane}, {Gonneau}, {Francois}, {Lewis}, {Sacco}, {Bragaglia}, {Heiter}, {Feltzing}, {Bensby}, {Irwin}, {Gonzalez Solares}, {Murphy}, {Bayo}, {Sbordone}, {Zwitter}, {Lanzafame}, {Walton}, {Zaggia}, {Alfaro}, {Morbidelli}, {Sousa}, {Monaco}, {Carraro}, \& {Lardo}}]{2020Worley}
{Worley}, C.~C., {Jofr{\'e}}, P., {Rendle}, B., {et~al.} 2020, \aap, 643, A83, \dodoi{10.1051/0004-6361/201936726}

\bibitem[{{Yan} {et~al.}(2022){Yan}, {Li}, {Wang}, {Zong}, {Yuan}, {Xiang}, {Huang}, {Xie}, {Dong}, {Yuan}, {Bi}, {Chu}, {Cui}, {Deng}, {Fu}, {Han}, {Hou}, {Li}, {Liu}, {Liu}, {Liu}, {Luo}, {Shi}, {Wu}, {Zhang}, {Zhao}, \& {Zhao}}]{2022Yan}
{Yan}, H., {Li}, H., {Wang}, S., {et~al.} 2022, The Innovation, 3, 100224, \dodoi{10.1016/j.xinn.2022.100224}

\bibitem[{{Yu} {et~al.}(2020){Yu}, {Bedding}, {Stello}, {Huber}, {Compton}, {Gizon}, \& {Hekker}}]{2020Yu}
{Yu}, J., {Bedding}, T.~R., {Stello}, D., {et~al.} 2020, \mnras, 493, 1388, \dodoi{10.1093/mnras/staa300}

\bibitem[{{Zhang} {et~al.}(2021){Zhang}, {Li}, {Yang}, {Xiong}, {Fu}, {Liu}, {Tian}, {Li}, {Wang}, {Liang}, {Zhou}, {Zong}, {Yang}, {Liu}, \& {Hou}}]{2021Zhang}
{Zhang}, B., {Li}, J., {Yang}, F., {et~al.} 2021, \apjs, 256, 14, \dodoi{10.3847/1538-4365/ac0834}

\bibitem[{{Zhang} {et~al.}(2022){Zhang}, {Jing}, {Yang}, {Wan}, {Ji}, {Fu}, {Liu}, {Zhang}, {Luo}, {Tian}, {Zhou}, {Wang}, {Guo}, {Zong}, {Xiong}, \& {Li}}]{2022Zhang}
{Zhang}, B., {Jing}, Y.-J., {Yang}, F., {et~al.} 2022, \apjs, 258, 26, \dodoi{10.3847/1538-4365/ac42d1}

\bibitem[{{Zhao} {et~al.}(2012){Zhao}, {Zhao}, {Chu}, {Jing}, \& {Deng}}]{2012Zhao}
{Zhao}, G., {Zhao}, Y.-H., {Chu}, Y.-Q., {Jing}, Y.-P., \& {Deng}, L.-C. 2012, Research in Astronomy and Astrophysics, 12, 723, \dodoi{10.1088/1674-4527/12/7/002}

\bibitem[{{Zong} {et~al.}(2024){Zong}, {Fu}, {Su}, {Hu}, {Zhang}, {Wang}, {Liu}, {Meng}, {Catanzaro}, {Frasca}, {Wang}, \& {Zong}}]{2024Zong}
{Zong}, P., {Fu}, J.-N., {Su}, J., {et~al.} 2024, \aj, 167, 227, \dodoi{10.3847/1538-3881/ad3357}

\bibitem[{{Zong} {et~al.}(2016){Zong}, {Charpinet}, {Vauclair}, {Giammichele}, \& {Van Grootel}}]{2016Zong}
{Zong}, W., {Charpinet}, S., {Vauclair}, G., {Giammichele}, N., \& {Van Grootel}, V. 2016, \aap, 585, A22, \dodoi{10.1051/0004-6361/201526300}

\bibitem[{{Zong} {et~al.}(2018){Zong}, {Fu}, {De Cat}, {Shi}, {Luo}, {Zhang}, {Frasca}, {Corbally}, {Molenda-{\.Z}akowicz}, {Catanzaro}, {Gray}, {Wang}, {Pan}, {Ren}, {Zhang}, {Jin}, {Wu}, {Dong}, {Xie}, {Zhang}, {Hou}, \& {LAMOST-Kepler Collaboration}}]{2018Zong}
{Zong}, W., {Fu}, J.-N., {De Cat}, P., {et~al.} 2018, \apjs, 238, 30, \dodoi{10.3847/1538-4365/aadf81}

\bibitem[{{Zong} {et~al.}(2020){Zong}, {Fu}, {De Cat}, {Wang}, {Shi}, {Luo}, {Zhang}, {Frasca}, {Molenda-{\.Z}akowicz}, {Gray}, {Corbally}, {Catanzaro}, {Cang}, {Wang}, {Chen}, {Hou}, {Liu}, {Niu}, {Pan}, {Tian}, {Yan}, {Zhang}, \& {Zuo}}]{2020Zong}
---. 2020, \apjs, 251, 15, \dodoi{10.3847/1538-4365/abbb2d}

\bibitem[{{Zuo} {et~al.}(2024){Zuo}, {Luo}, {Du}, {Li}, {Jones}, {Song}, {Kong}, \& {Guo}}]{2024Zuo}
{Zuo}, F., {Luo}, A.~L., {Du}, B., {et~al.} 2024, \apjs, 271, 4, \dodoi{10.3847/1538-4365/ad1eeb}

\end{thebibliography}
\bibliographystyle{aasjournal}

\appendix

\section{Parameters of melta-poor star candidates}\label{melta_poor_tab}

\begin{table*}[ht]
\centering
\caption{Metal-poor star candidates identified in LK-MRS--\uppercase\expandafter{\romannumeral1}.}
\label{tab:metal_poor}
\begin{threeparttable}
\begin{tabular*}{\textwidth}{@{\extracolsep{\fill}} c c c c r r r c c}
\toprule
Times & uid & \textit{Gaia} Source & KIC/EPIC & RA (deg) & Dec (deg) & [Fe/H] (dex) & Label & Discovery \\
\midrule
1 & G16299287415772 & 598414739630537600  & 211344111  & 130.74812 & 10.70665 & $-1.33$ & MP  & Known \\
1 & G16343607386694 & 605159247034372224  & --        & 133.29913 & 12.93157 & $-2.45$ & VMP & New   \\
1 & G15632253923687 & 3865499501595685376 & 248633859  & 161.03179 & 6.72897  & $-1.10$ & MP  & New   \\
1 & G14076948140108 & 2080028820098431616 & --        & 295.39068 & 46.07732 & $-2.05$ & VMP & New   \\
2 & G16330175937848 & 689476876759791360  & 212170836  & 132.26083 & 23.50690 & $-1.07$ & MP  & New   \\
1 & G16713619459999 & 144889557795806848  & --        & 69.24472  & 21.74420 & $-2.36$ & VMP & New   \\
1 & G16329236204097 & 652853175231968128  & --        & 125.90789 & 15.91996 & $-1.11$ & MP  & New   \\
1 & G17073665209151 & 2538128245175162496 & --        & 16.52815  & 1.37751  & $-2.18$ & VMP & New   \\
1 & G16585545411581 & 147111567716180096  & 247808458  & 72.65588  & 24.46694 & $-2.23$ & VMP & New   \\
\bottomrule
\end{tabular*}
\begin{tablenotes}
\item[] Column\,1 provides the number of available spectral parameters for each target. Column\,2 presents the uid, while Column\,3 gives the \textit{Gaia} Source ID. Column\,4 provides the KIC/EPIC identifier (if available; "--" denotes no match). Columns\,5 and 6 contain the RA and Dec in J2000.0 coordinates. Column\,7 presents the [Fe/H] metallicity in dex. Column\,8 provides the metallicity classification (MP = metal-poor with $-2.0 \lesssim \mathrm{[Fe/H]} < -1.0$; VMP = very metal-poor with $\mathrm{[Fe/H]} \lesssim -2.0$). Column\,9 indicates the discovery status (Known = present in \textsc{SIMBAD}; New = newly recognized in this work). The complete table will be available online in a machine-readable format.
\end{tablenotes}
\end{threeparttable}
\end{table*}

\section{Parameters of high velocity star candidates}\label{high_velocity_tab}
\begin{table*}[ht]
\centering
\small
\caption{High velocity star candidates identified in LK-MRS–\uppercase\expandafter{\romannumeral1}.}
\label{tab:high_velocity}
\begin{threeparttable}
\begin{tabular*}{\textwidth}{@{\extracolsep{\fill}} c c c r r r r c c}
\toprule
Times & uid & KIC/EPIC & RA (deg) & Dec (deg) & RV (km/s) & $V_{\mathrm{gc}}$ (km/s) & Label & Discovery \\
\midrule
1  & G14078689694621 & 10796857   & 291.68157 & 48.12203  & 132.45   & 374.02   & bound    & New   \\
4  & G17525854519286 & —          & 55.56701  & 25.26853  & 159.08   & 354.31   & bound    & New   \\
4  & G17525829022163 & —          & 55.93135  & 25.54486  & 261.31   & 356.68   & bound    & New   \\
5  & G14087386039950 & 7612547    & 293.37042 & 43.25230  & 169.08   & 434.35   & bound    & New   \\
12 & L15619442062806 & —          & 158.64312 & 5.56515   & -37.87   & 360.64   & bound    & Known \\
8  & G17065374200707 & 220496220  & 13.10688  & 6.68008   & 215.77   & 374.40   & unbound  & Known \\
5  & G14081832652116 & 5953450    & 289.59470 & 41.23543  & -0.02    & 394.47   & bound    & New   \\
3  & G15626569425898 & 248671171  & 157.98600 & 7.64722   & -34.70   & 369.90   & bound    & New   \\
2  & G17075733875360 & —          & 16.36599  & 3.55150   & 96.19    & 351.40   & bound    & New   \\
1  & G16295662579783 & 211440176  & 129.72939 & 12.24899  & -258.68  & 397.15   & bound    & New   \\
\bottomrule
\end{tabular*}
\begin{tablenotes}
\item Columns\,1–3 list the number of available spectral parameters for each target, uid, and KIC/EPIC ID (if available). Columns\,4–5 give J2000 coordinates. Columns\,6–7 list heliocentric radial velocity and the Galactic standard-of-rest velocity. Column\,8 gives the kinematic label, and column\,9 indicates whether the star was previously known or newly identified. The full table, including Gaia source IDs, is available online.
\end{tablenotes}
\end{threeparttable}
\end{table*}

\section{Parameters of RV variable star candidates}\label{rvv_tab}

\begin{table*}[ht]
\centering
\small
\caption{RV variable star candidates of the LK-MRS-\uppercase\expandafter{\romannumeral1}.}
\label{tab:rv_variable}
\begin{threeparttable}
\begin{tabular*}{\textwidth}{@{\extracolsep{\fill}} c r r c r r r c c c c}
\toprule
uid & RA & Dec & MJD & S/N & RV & $RV_{\mathrm{error}}$ & KIC/EPIC & TIC & Label & Discovery \\
 & (deg) & (deg) & (day) &  & (km/s) & (km/s) &  &  &  & \\
\midrule
G14075814741395 & 288.904419 & 44.520668 & 58624 & 70.17  & -49.97 & 1.13 & —          & 158844295 & gamma & Known \\
G14075814741395 & 288.904419 & 44.520668 & 58645 & 65.61  & -51.98 & 1.18 & —          & 158844295 & gamma & Known \\
G14075814741395 & 288.904419 & 44.520668 & 58269 & 40.98  & -37.12 & 0.62 & —          & 158844295 & gamma & Known \\
G14075814741395 & 288.904419 & 44.520668 & 58266 & 77.94  & -37.27 & 0.66 & —          & 158844295 & gamma & Known \\
G14075814741395 & 288.904419 & 44.520668 & 58643 & 120.28 & -53.10 & 1.18 & —          & 158844295 & gamma & Known \\
L17526106473874 &  57.878652 & 25.140276 & 58829 & 57.86  & -25.75 & 1.41 & 211134060.0 &  84331634 & gamma & New   \\
L17526106473874 &  57.878652 & 25.140276 & 58835 & 18.90  & -21.54 & 1.39 & 211134060.0 &  84331634 & gamma & New   \\
L17526106473874 &  57.878652 & 25.140276 & 59180 & 42.79  & -14.59 & 1.32 & 211134060.0 &  84331634 & gamma & New   \\
L17526106473874 &  57.878652 & 25.140276 & 59189 & 85.94  & -13.35 & 1.33 & 211134060.0 &  84331634 & gamma & New   \\
L17526106473874 &  57.878652 & 25.140276 & 58851 & 126.35 & -20.06 & 1.31 & 211134060.0 &  84331634 & gamma & New   \\
\bottomrule
\end{tabular*}
\begin{tablenotes}
\item Column\,1 provides the uid for each target. Columns\,2 and\,3 list the RA and Dec in J2000.0 coordinates. Column\,4 contains the modified Julian date (MJD) of the observations. Column\,5 gives the S/N. Columns\,6 and\,7 provide the measured RV and its associated uncertainty . Columns\,8 and\,9 list the KIC/EPIC and TIC IDs (if available). Column\,10 indicates the variability class, and Column\,11 specifies the discovery status. The complete table, including Gaia source IDs, will be available online in a machine-readable format.
\end{tablenotes}
\end{threeparttable}
\end{table*}

\end{document}